\documentclass[aps,prx,superscriptaddress,amsfonts,amsmath,amssymb,twocolumn,floatfix,reprint]{revtex4-1}

\usepackage{bm}
\usepackage{graphicx}
\usepackage{amsmath}
\usepackage{amstext}
\usepackage{amssymb}
\usepackage{amsfonts}
\usepackage{amsbsy}
\usepackage{verbatim}
\usepackage[usenames,dvipsnames,svgnames,table]{xcolor}
\usepackage{multirow}
\usepackage{gensymb}
\usepackage{textcomp}
\usepackage{enumitem}
\usepackage{comment}
\usepackage{fancyhdr}
\usepackage[colorlinks=true, urlcolor=Blue, linkcolor=Blue, citecolor=Blue, pdftex]{hyperref}
\usepackage{times}
     
\begin{document}

\title{Persistence of the gapless spin liquid in the breathing kagome Heisenberg antiferromagnet}

\author{Yasir Iqbal}
\email[]{yiqbal@physics.iitm.ac.in}
\affiliation{Department of Physics, Indian Institute of Technology Madras, Chennai  600036, India}
\author{Didier Poilblanc}
\affiliation{Laboratoire de Physique Th\'eorique UMR-5152, CNRS and Universit\'e de Toulouse, F-31062, 
Toulouse, France}
\author{Ronny Thomale}
\affiliation{Institute for Theoretical Physics and Astrophysics, Julius-Maximilian's
University of W\"urzburg, Am Hubland, D-97074 W\"urzburg, Germany}
\author{Federico Becca}
\affiliation{Democritos National Simulation Center, Istituto Officina dei Materiali del CNR and 
SISSA-International School for Advanced Studies, Via Bonomea 265, I-34136 Trieste, Italy}

\date{\today}

\begin{abstract}
The nature of the ground state of the spin $S=1/2$ Heisenberg antiferromagnet on the kagome lattice with breathing anisotropy (i.e., with 
different superexchange couplings $J_{\vartriangle}$ and $J_{\triangledown}$ within elementary up- and down-pointing triangles) is investigated 
within the framework of Gutzwiller projected fermionic wave functions and Monte Carlo methods. We analyze the stability of the U(1) Dirac spin 
liquid with respect to the presence of fermionic pairing that leads to a gapped $\mathbb{Z}_{2}$ spin liquid. For several values of the ratio 
$J_{\triangledown}/J_{\vartriangle}$, the size scaling of the energy gain due to the pairing fields and the variational parameters are reported. 
Our results show that the energy gain of the gapped spin liquid with respect to the gapless state either vanishes for large enough system size 
or scales to zero in the thermodynamic limit. Similarly, the optimized pairing amplitudes (responsible for opening the spin gap) are shown to 
vanish in the thermodynamic limit. Our outcome is corroborated by the application of one and two Lanczos steps to the gapless and gapped wave 
functions, for which no energy gain of the gapped state is detected when improving the quality of the variational states. Finally, we discuss 
the competition with the ``simplex" $\mathbb{Z}_{2}$ resonating-valence-bond spin liquid, valence-bond crystal, and nematic states in the strongly 
anisotropic regime, i.e., $J_{\triangledown} \ll J_{\vartriangle}$.
\end{abstract}

\maketitle

\section{Introduction}

In the past two decades, considerable effort has been devoted towards understanding the properties of the $S=1/2$ Heisenberg model on the kagome 
lattice, which represents the purest example of geometric frustration in two dimensions. This is reflected in the fact that the ground state fails 
to develop long-range magnetic order, thus potentially realizing a {\it quantum spin liquid} phase~\cite{Pomeranchuk1941}, which features high 
entanglement, low-energy excitations with fractional quantum numbers, and possibly topological order~\cite{Balents2010,Savary2017,Zhou2017}. Even 
though investigations of the Heisenberg model on the kagome lattice started in the 1990s~\cite{Zeng1990,Sachdev1992,Lecheminant1997}, a considerable 
boost was given by the discovery of Herbertsmithite [ZnCu$_3$(OH)$_6$Cl$_2$], which proves to be an excellent embodiment of the nearest-neighbor
$S=1/2$ Heisenberg model on the structurally perfect kagome lattice, with only minor longer-range super-exchange couplings~\cite{Shores2005,Mendels2007,Helton2007,Suttner2014}. 
Experimental investigations have revealed the absence of long-range magnetic order or frozen magnetic moments; however, in the resulting quantum 
spin liquid, it has been particularly challenging to reach a definite conclusion as to the presence/absence of a spin gap in the excitation 
spectrum which is expected to be tiny~\cite{Olariu2008,Han2012,Fu2015}. Similarly, theoretical approaches have long wrestled with the question of 
the nature of the ground state and properties of its low-energy excitations, which turn out to be particularly elusive and remain perplexing. 
Indeed, early density-matrix renormalization group (DMRG) calculations reported the presence of a finite $S=1$ gap~\cite{Jiang2008,Yan2011}, 
suggestive of a topologically ordered $\mathbb{Z}_{2}$ spin liquid ground state~\cite{Depenbrock2012}. In contrast, recent calculations based 
upon Gutzwiller projected fermionic wave functions~\cite{Ran2007,Iqbal2013,Iqbal2014,Iqbal2015}, DMRG~\cite{He2017,Zhu-2018}, and tensor network 
approaches~\cite{Liao2017} provide strong evidence in favor of a gapless spin liquid with signatures of Dirac cones in the spinon spectrum.

\begin{figure}
\includegraphics[width=0.9\columnwidth]{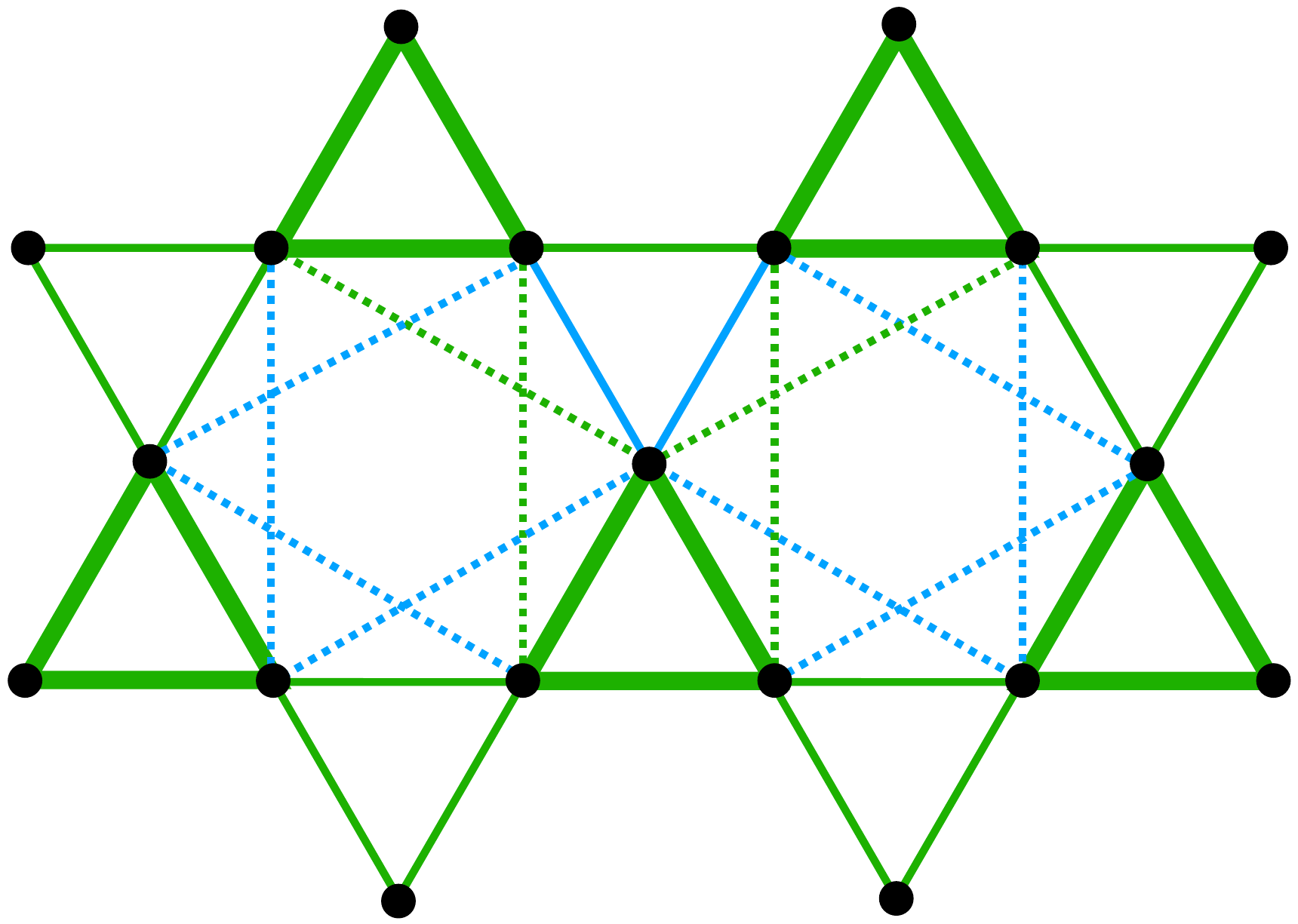}
\caption{\label{fig:Ansatz} 
Breathing kagome lattice is defined with nearest-neighbor superexchange coupling $J_{\vartriangle}$ on up-pointing triangles (thick solid 
lines) and $J_{\triangledown}$ on down-pointing triangles (thin solid lines). A schematic illustration of the $\mathbb{Z}_{2}[0,\pi]\beta^{*}$ 
spin liquid Ansatz (Ref.~\cite{Schaffer2017}) is also shown. The auxiliary (spinon) Hamiltonian requires a $2 {\times} 1$ doubling of the 
three-site geometrical unit cell. Nearest-neighbor (next-nearest-neighbor) bonds are shown by solid (dashed) lines. The green (blue) bonds represent
$s_{ij}=\nu_{ij}=+1$ ($s_{ij}=\nu_{ij}=-1$) in Eq.~(\ref{eq:MF-Z2}). The fact that the hopping and pairing amplitudes on nearest-neighbor 
bonds belonging to up- and down-pointing triangles are allowed to be different is represented by a difference in the  thickness of bonds.}
\end{figure}

In order to reach a consensus on the low-energy properties of the $S=1/2$ Heisenberg model on the kagome lattice, and the possibility of 
it describing the experimental features observed in ZnCu$_3$(OH)$_6$Cl$_2$, it proves enlightening to look at variations of the model arising 
from distortions of the geometrically perfect kagome lattice. On a more conceptual level, it is a recurrent motif in theoretical physics to 
introduce interpolation parameters in order to facilitate the model analysis of a particular parameter limit. As it has been suggested early 
on that the nearest neighbor $S=1/2$ Heisenberg model might be located close to a first order phase transition, it appears useful to introduce a 
geometric distortion parameter, and study the model family as it approaches the isotropic limit. Concretely, it may offer an alternative 
route for the study of quantum spin liquids. One example is given by Volborthite [Cu$_3$V$_2$O$_7$(OH)$_2 \cdot 2$H$_2$O], where the elementary 
triangles that build up the kagome lattice are no longer equilateral but isosceles, leading to different antiferromagnetic couplings along short 
and long bonds. In this case, there is some evidence for a magnetic ground state, even though unusually slow spin fluctuations persist down to 
low temperatures~\cite{Bert2005,Yavorskii2007,Wang2007,Yoshida2009,Janson2010,Nilsen2011,Janson2016,Chern2017a,Chern2017b}. Another interesting 
deformation is one leading to alternately sized equilateral triangles, dubbed the {\it trimerized} or {\it breathing} kagome lattice~\cite{Aidoudi2011}, 
in analogy to the breathing pyrochlores~\cite{Okamoto2013}. Correspondingly, the kagome lattice features an alternation of interactions, with 
the triangles pointing up (having a superexchange coupling $J_{\vartriangle}$) and those pointing down (with $J_{\triangledown}$)~\cite{Essafi2017}; 
see Fig.~\ref{fig:Ansatz}. Originally, this model was considered by Mila~\cite{Mila1998,Mambrini2000}, in order to explain 
the large number of singlet excitations at low energies detected within exact diagonalizations on small clusters for the isotropic limit with 
$J_{\vartriangle}=J_{\triangledown}$~\cite{Lecheminant1997}. Remarkably, vanadium oxyfluoride (NH$_4$)$_2$[C$_7$H$_{14}$N][V$_7$O$_6$F$_{18}$] 
(DQVOF) provides a realization of the breathing kagome lattice with $J_{\triangledown}/J_{\vartriangle}=0.55(4)$~\cite{Orain2017}. An earlier 
muon spin resonance ($\mu$SR) study~\cite{Orain2014} and a more recent nuclear magnetic resonance (NMR) study revealed no magnetic order, with 
the latter pointing to an essentially gapless excitation spectrum~\cite{Clark2013,Orain2017}. These results have provided a renewed impetus to 
understand whether a gapless spin liquid may be stabilized in realistic spin models with SU(2) symmetry.

The Hamiltonian for the breathing kagome lattice is given by
\begin{equation}\label{eq:model}
{\cal \hat{H}} = J_{\vartriangle} \sum_{\langle ij\rangle\in\vartriangle} {\bf \hat{S}}_{i} \cdot {\bf \hat{S}}_{j} +
J_{\triangledown} \sum_{\langle ij\rangle\in\triangledown} {\bf \hat{S}}_{i} \cdot {\bf \hat{S}}_{j}, 
\end{equation}
where ${\bf \hat{S}}_{i}=(\hat{S}_i^x,\hat{S}_i^y,\hat{S}_i^z)$ is the $S=1/2$ operator on a site $i$ and $\langle ij\rangle$ indicate 
nearest-neighbor pairs of sites $i$ and $j$ that belong to up-pointing ($\langle ij\rangle \in$ $\vartriangle$) or down-pointing 
($\langle ij\rangle\in\triangledown$) triangles. The crystallographic unit cell of this lattice consists of three sites located at $(0,0)$, 
$(1,0)$, and $(1/2,\sqrt{3}/2)$ (forming an up-pointing triangle); the primitive vectors are ${\bf a}_1=(2,0)$ and ${\bf a}_2=(1,\sqrt{3})$.
For our calculations, we consider toric clusters that are defined by ${\bf T}_1=L {\bf a}_1$ and ${\bf T}_2=L {\bf a}_2$, and thus consist
of $N=3 L^2$ sites. Notice that, for $J_{\triangledown}=0$ (or $J_{\vartriangle}=0$) the Hamiltonian corresponds to uncoupled up-pointing
(or down-pointing) triangles. At this special point, the ground state is highly degenerate, since each interacting triangle has a doubly degenerate 
ground state with an energy per triangle $E_{\vartriangle}=-3/4J_{\vartriangle}$ ($E_{\triangledown}=-3/4J_{\triangledown}$) and spin $S=1/2$. 
In the weakly coupled limits $J_{\triangledown} \ll J_{\vartriangle}$ (or $J_{\vartriangle} \ll J_{\triangledown}$), the massive degeneracy is 
expected to be partially or completely lifted. A perturbative treatment around the uncoupled limit, unfortunately, gives rise to a complicated 
effective model~\cite{Mila1998}, which contains both spin and pseudospin degrees of freedom and whose solution cannot be obtained in a 
straightforward manner.

Recently, the Heisenberg model on the breathing kagome lattice has been investigated theoretically by using a projective-symmetry group (PSG) 
analysis supplemented by Monte Carlo simulations of variational wave functions~\cite{Schaffer2017} and by DMRG calculations~\cite{Repellin2017}.
The latter one pointed to the existence of an extended gapless spin liquid phase which shows signatures of Dirac cones, similar to what has 
been found at the isotropic point~\cite{He2017}. In the limit of strong breathing anisotropy $J_{\triangledown}\ll J_{\vartriangle}$, the 
existence of a lattice-nematic state, i.e., a state with inequivalent nearest-neighbor spin-spin correlations, was claimed for in the regime 
$J_{\triangledown}/J_{\vartriangle} \lesssim 0.13$. In contrast, the variational Monte Carlo study claimed that a gapped $\mathbb{Z}_{2}$ 
spin liquid ground state is obtained within Gutzwiller projected fermionic wave functions. However, this conclusion was based only upon a 
calculation of the variational parameters and energies for a few system sizes without a finite-size-scaling analysis. 

In this paper, we report a high-accuracy systematic study of both the U(1) Dirac state and the gapped $\mathbb{Z}_{2}$ state that is obtained
from the U(1) Dirac state by an inclusion of a fermionic pairing term. By performing calculations on very large system sizes (up to $N=2352$ 
sites), we show that the variational parameters that are responsible for a finite spin gap are vanishing in the thermodynamic limit and, therefore, 
the energy gain of the gapped $\mathbb{Z}_{2}$ state with respect to the U(1) Dirac state scales to zero for $N \to \infty$. Moreover, in the 
strongly anisotropic limit $J_{\triangledown}\ll J_{\vartriangle}$, we show that the U(1) Dirac spin liquid undergoes a dimer instability, 
giving way to a valence-bond crystal (VBC) ground state for $0<J_{\triangledown}/J_{\vartriangle} \lesssim 0.25$. In addition, in this regime,
a ``simplex'' $\mathbb{Z}_{2}$ resonating-valence-bond (RVB) spin liquid is found to have an energy between the U(1) Dirac state and the VBC 
state. 

The paper is organized as follows: in Sec.~\ref{sec:wavefunction}, we describe the variational wave functions that are used in this work 
(and also the simplex RVB state that is constructed and used within a tensor-network approach); in Sec.~\ref{sec:results}, we present our
results; finally, in Sec.~\ref{sec:concl}, we draw our conclusions.

\section{Variational wave functions}\label{sec:wavefunction}

\subsection{Gutzwiller projected Ans\"atze}

The variational wave functions are written in terms of Abrikosov fermions~\cite{Abrikosov1965}. In the following, the noninteracting state, 
defined in the fermionic Hilbert space, is obtained by taking the ground state $|\Phi_0\rangle$ of the following auxiliary Hamiltonian, which
has the form of a generalized Bardeen-Cooper-Schrieffer (BCS) Hamiltonian:
\begin{eqnarray}
&&{\cal \hat{H}}_{{\rm aux}}\{\mathbb{Z}_{2}[0,\pi]\beta^{*}\} =
\chi_{\vartriangle}\sum_{\langle ij\rangle\in\vartriangle,\alpha}{\rm s}_{ij} \hat{c}_{i,\alpha}^{\dagger}\hat{c}_{j,\alpha} \nonumber \\&+&
\sum_{\langle ij\rangle\in\triangledown}{\rm s}_{ij}\{\chi_{\triangledown}\sum_{\alpha}\hat{c}^{\dagger}_{i,\alpha}\hat{c}_{j,\alpha}+
\Delta_{\triangledown}(\hat{c}^{\dagger}_{i,\uparrow}\hat{c}^{\dagger}_{j,\downarrow}+ {\rm h.c.})\}\nonumber \\
&+& \sum_{\langle\langle ij\rangle\rangle}{\nu}_{ij}\{\chi_{2}\sum_{\alpha}\hat{c}^{\dagger}_{i,\alpha}\hat{c}_{j,\alpha}+
\Delta_{2}(\hat{c}^{\dagger}_{i,\uparrow}\hat{c}^{\dagger}_{j,\downarrow}+ {\rm h.c.})\}\nonumber \\
&+&
\sum_{i}\{\mu\sum_{\alpha}\hat{c}_{i,\alpha}^{\dagger}\hat{c}_{i,\alpha}
+\zeta (\hat{c}_{i,\uparrow}^{\dagger}\hat{c}_{i,\downarrow}^{\dagger}+{\rm h.c.})\};
\label{eq:MF-Z2}
\end{eqnarray}
Here, $\langle ij\rangle\in{\vartriangle}$ and $\langle ij\rangle\in{\triangledown}$ denote sums over pairs of nearest-neighbor sites belonging
to up- and down-pointing triangles, respectively, while $\langle\langle ij\rangle\rangle$ denote sums over pairs of next-nearest-neighbor sites; 
${\rm s}_{ij}$ and $\nu_{ij}$ encode the sign structure of the nearest- and next-nearest-neighbor pairs of sites, as depicted in 
Fig.~\ref{fig:Ansatz}. The variational wave function thus obtained contains six variational parameters (upon fixing $\chi_{\vartriangle}=1$ as 
the overall energy scale), namely, the nearest-neighbor hopping ($\chi_{\triangledown}$) and pairing ($\Delta_{\triangledown}$) on down-pointing 
triangles, the next-nearest-neighbor hopping ($\chi_{2}$) and pairing ($\Delta_{2}$), the onsite chemical potential ($\mu$), and real on-site 
pairing ($\zeta$). In order to have a nondegenerate ground state of the auxiliary Hamiltonian, we choose antiperiodic and periodic boundary 
conditions along ${\bf a}_1$ and ${\bf a}_2$, respectively.

The form of this Ansatz is dictated by the PSG classification~\cite{Schaffer2017}, and it describes both the gapless U(1) Dirac state (when all 
the fermionic pairing terms $\Delta_{\triangledown}$, $\Delta_{2}$, and $\zeta$ are identically zero) and a generalization of the so-called
$\mathbb{Z}_{2}[0,\pi]\beta$ state that was obtained for the isotropic limit~\cite{Lu2011} (when at least one pairing amplitude is nonzero), 
and hereafter is referred to as the $\mathbb{Z}_{2}[0,\pi]\beta^{*}$ spin liquid. In total, the PSG approach for the breathing kagome lattice allows 
for six different $\mathbb{Z}_{2}$ Ans\"atze~\cite{Schaffer2017}. However, two of them do not allow any amplitudes on nearest-neighbor pairs of 
sites nor any on-site (chemical potential and pairing) terms, thus making the variational Ansatz unplausible for a model with $J_{\vartriangle} \ne 0$
and $J_{\triangledown} \ne 0$; for another two Ans\"atze, the on site and nearest-neighbor pairings are not allowed, which again renders them 
energetically unfavorable; finally, among the remaining two options, one has the uniform flux structure with ${\rm s}_{ij}=\nu_{ij}=+1$, which 
gives a rather high variational energy, while the last one (the $\mathbb{Z}_{2}[0,\pi]\beta^{*}$ spin liquid) is parametrized by the Hamiltonian 
of Eq.~(\ref{eq:MF-Z2}). 

\begin{figure}
\includegraphics[width=0.85\columnwidth]{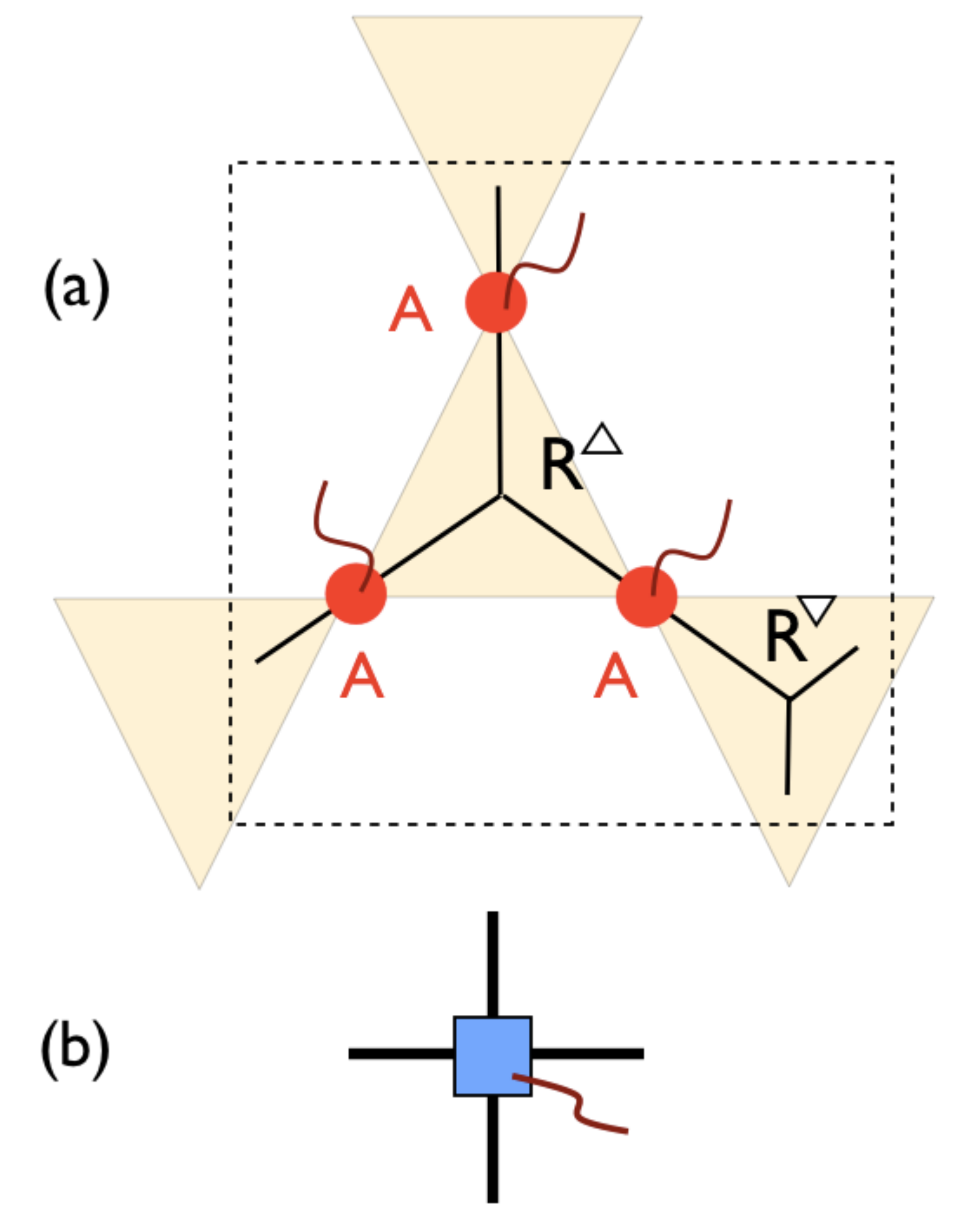}
\caption{\label{fig:PEPS}
(a) Local tensors defining the RVB PEPS on the kagome lattice. Straight (wiggly) lines denote virtual (physical) degrees of freedom, spanning a 
Hilbert space of dimension $D=3$ ($d=2$). In the simplex RVB, one applies the operator $\mathbb{I}-\alpha \mathbb{P}_{3/2}$ on the three wiggly 
lines. (b) After grouping the three sites of the up triangle, one obtains a rank-5 four-coordinated tensor.}
\end{figure}

A \emph{bona fide} spin liquid wave function, which lives in the correct Hilbert space with one fermion per site (corresponding to the physical 
Hilbert space of the spin model), is obtained by applying the Gutzwiller projector to the noninteracting state $|\Phi_0\rangle$:
\begin{equation}
|\Psi_{{\rm SL}}\rangle = {\cal P}_{\text G} |\Phi_0\rangle,
\end{equation}
where ${\cal P}_{\text G}= \prod_i \left ( \hat{n}_{i,\uparrow}-\hat{n}_{i,\downarrow} \right )$, $\hat{n}_{i,\alpha} = \hat{c}_{i,\alpha}^{\dagger}\hat{c}_{i,\alpha}$ 
being the fermionic density per spin $\alpha$ on the site $i$. The variational energy and correlation functions over $|\Psi_{{\rm SL}}\rangle$ can 
be calculated in a straightforward manner by using Monte Carlo sampling~\cite{BeccaBook}. In addition, a stochastic optimization is possible to 
obtain accurate estimations of the variational parameters contained in Eq.~(\ref{eq:MF-Z2})~\cite{BeccaBook,Sorella2005}.

We would like to mention that the Gutzwiller projected wave function, with only $\chi_{\vartriangle}=1$ (or $\chi_{\triangledown}=1$) and all the other parameters equal to zero, 
gives the exact energy in the limit of decoupled triangles with $J_{\triangledown}=0$ (or $J_{\vartriangle}=0$) and represents, in the general case,
an excellent approximation for the isotropic case with $J_{\triangledown}=J_{\vartriangle}$~\cite{Iqbal2011b}.

The accuracy of the variational wave functions can be easily improved by applying a few Lanczos steps on the variational state~\cite{Becca2015}:
\begin{equation}\label{eq:lanczos}
|\Psi_{p{-}\rm{LS}}\rangle =  \left ( 1+\sum_{k=1}^{p}\alpha_{k}{\cal \hat{H}}^{k} \right )|\Psi_{{\rm SL}}\rangle,
\end{equation}
where $\{ \alpha_{k} \}$ is a set of variational parameters. On large cluster sizes, only a few steps can be efficiently implemented, and 
here we consider the case with $p=1$ and $p=2$ ($p=0$ corresponds to the original trial wave function). In addition, an estimate of the 
exact ground-state energy may be achieved by the method of variance extrapolation. In fact, for sufficiently accurate states, we have that
$E-E_{\rm ex} \approx \sigma^{2}$, where $E=\langle {\cal \hat{H}} \rangle/N$ and 
$\sigma^{2}=(\langle {\cal \hat{H}}^{2}\rangle{-}\langle {\cal \hat{H}}\rangle^{2})/N$ are the energy and variance per site, respectively; 
therefore, the exact ground-state energy $E_{\rm ex}$ can be extracted by fitting $E$ vs $\sigma^{2}$ for $p=0$, $1$, and $2$. Also, in the presence 
of a few Lanczos steps the energy and its variance can be obtained using the standard variational Monte Carlo method.

\begin{figure*}
\includegraphics[width=\textwidth]{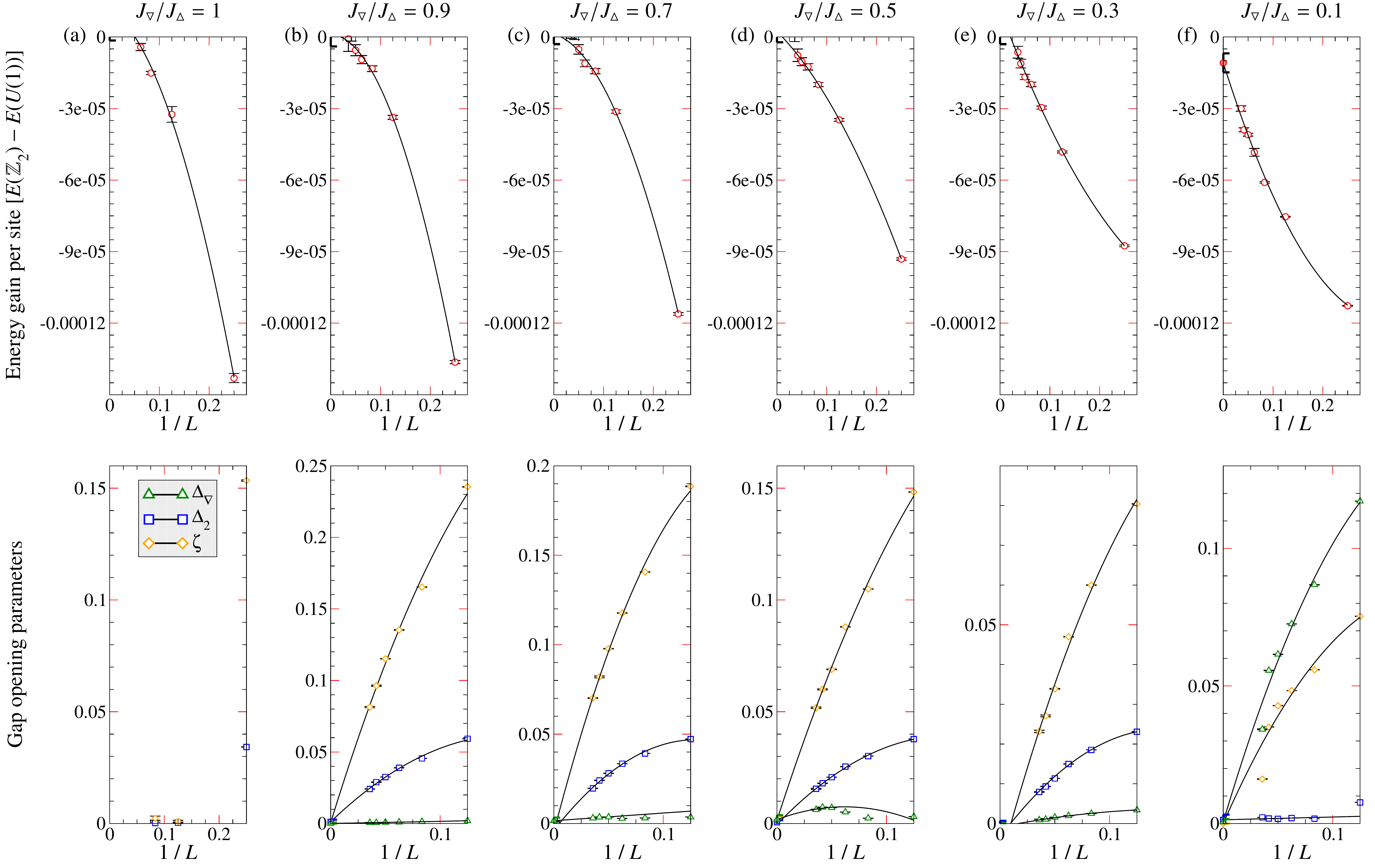}
\caption{\label{fig:energy-gain}
For different values of the breathing anisotropy $J_{\triangledown}/J_{\vartriangle}$, we show the finite-size scaling of the energy gain of 
the $\mathbb{Z}_{2}[0,\pi]\beta^{*}$ spin liquid with respect to the U(1) Dirac spin liquid, i.e., $E(\mathbb{Z}_{2})-E(\rm{U}(1))$ (first row). 
The finite-size scaling of $\Delta_{\triangledown},~\Delta_{2},~\zeta$, the variational parameters responsible for opening a gap, are also 
shown (second row). Here, lines are quadratic fits~\cite{extr} of the results. The largest cluster considered corresponds to $L=28$ and has $2352$ sites. 
The results for the isotropic limit $J_{\triangledown}/J_{\vartriangle}=1$ are also reported for comparison.}
\end{figure*}

\subsection{The simplex RVB as a Projected Entangled Pair State}

Other types of spin liquids can be constructed using the framework of projected-entangled pair states (PEPS)~\cite{Verstraete2004b,Schuch2007}. 
On a kagome lattice, a PEPS can be defined in terms of rank-$3$ tensors (i) $A^{s}_{\lambda,\mu}$ on the sites and (ii) 
$R^\vartriangle_{\lambda,\mu,\nu}$ and $R^\triangledown_{\lambda,\mu,\nu}$ in the center of the up- and down-pointing triangles, respectively, 
where $s=0,1$ are qubits representing the two $S_z=\pm 1/2$ spin components and $\lambda,\mu,\nu\in \{0,1,\dots,$D$\}$ are virtual indices, as 
shown in Fig.~\ref{fig:PEPS}(a)~\cite{Schuch2012}. One can then group three sites on each unit cell (for example, on the up-pointing triangles) to 
obtain a rank-$5$ tensor (of new physical dimension $2^3=8$) connected on an effective square lattice, as shown in Fig.~\ref{fig:PEPS}(b). 
The amplitudes of the PEPS in the local $S_z$ basis are then obtained by contracting all virtual indices. 

The original nearest-neighbor (NN) RVB state~\cite{Anderson1973} defined as an equal weight (and equal sign) summation of all NN singlet coverings
(NN  singlets are all oriented clockwise on all the triangles) also belongs to the class of short-ranged (topologically ordered) $\mathbb{Z}_2$
spin liquids. Such a state can in fact be represented as a PEPS with bond dimension $D=3$~\cite{Schuch2012,Poilblanc2012} and involving the above 
rank-$3$ tensors, $A^{s}_{\lambda,\mu}$ on the sites, and $R^\vartriangle_{\lambda,\mu,\nu}=R^\triangledown_{\lambda,\mu,\nu}=R_{\lambda,\mu,\nu}$
in the center of the triangles. More precisely, $A^{s}_{2,s}=A^{s}_{s,2}=1$, and zero otherwise, and $R_{2,2,2}=1$, and 
$R_{\lambda,\mu,\nu}=\epsilon_{\lambda,\mu,\nu}$ otherwise, with $\epsilon_{\lambda,\mu,\nu}$ being the antisymmetric tensor. Note that the RVB 
state is also equivalent to a projected BCS wave function~\cite{YangYao2012} and is perfectly (spatially) isotropic. It has been studied in 
detail in Ref.~\cite{Poilblanc2012} and its energy density was found to be rather poor compared to variational wave functions or DMRG.
In fact, the NN RVB wave function has a fixed proportion ($1/4$) of ``defect triangles'' with no singlet bonds (characterized by $\lambda=\mu=\nu=2$
on the three bonds of the corresponding PEPS $R$ tensor), equally distributed between the up- and down-pointing triangles. In the isotropic case 
$J_{\triangledown}=J_{\vartriangle}$, defect triangles are energetically costly. However, in the regime with strong anisotropy, i.e., 
$J_{\triangledown}\ll J_{\vartriangle}$, placing defects predominantly on the down-pointing triangles will be energetically very 
favorable~\cite{Mambrini2000}. Such an improvement can be performed easily within the PEPS formalism. Choosing the up-pointing triangles as the 
three-site units, one then acts with the operator $\mathbb{I}-\alpha \mathbb{P}_{3/2}$ on every unit (where $\mathbb{I}$ is the identity operator,
$\mathbb{P}_{3/2}$ is the projector on the fully symmetric subspace of three spins $1/2$, and $\alpha$ is a variational 
parameter~\cite{Poilblanc2013a}). As a result of this projection, we expect longer range singlet bonds to appear in the RVB state,
with a nontrivial sign structure.  When $\alpha=1$, one projects exactly onto the (two-dimensional) $S=1/2$ manifolds of all up-pointing triangles. 

\section{Results}\label{sec:results}

\subsection{Competition between the U(1) Dirac and gapped $\mathbb{Z}_{2}$ spin liquids}

Our main results are shown in Fig.~\ref{fig:energy-gain}. Here, we report the finite-size scaling of the on-site $\zeta$, nearest-neighbor 
$\Delta_{\triangledown}$, and next-nearest-neighbor $\Delta_2$ pairing terms for $J_{\triangledown}/J_{\vartriangle}=0.1$, $0.3$, $0.5$, 
$0.7$, and $0.9$; the isotropic case $J_{\triangledown}/J_{\vartriangle}=1$ is also reported for comparison. For all ratios of 
$J_{\triangledown}/J_{\vartriangle}$, we considered clusters for which $L=4n$ with $n$ ranging from $1$ to $7$ (the largest cluster thus has 
$N=2352$ sites), except for the isotropic point, where the maximum is $n=4$, since already for $n=3$ the pairing terms are vanishing. 
In addition, we also report the energy gain of the $\mathbb{Z}_{2}[0,\pi]\beta^{*}$ state due to the presence of these pairing variational 
parameters with respect to the U(1) Dirac state that contains only hopping terms, i.e., the gain $\Delta E = E(\mathbb{Z}_{2}) {-} E(\rm{U}(1))$ 
[see Supplemental Material (Ref.~\cite{supp}) for values of energies of the U(1) and $\mathbb{Z}_{2}[0,\pi]\beta^{*}$ spin liquids]. 

We find that for all values of $J_{\triangledown}/J_{\vartriangle}$ the pairing amplitudes scale to zero (within error bars) in the thermodynamic 
limit indicating that the $\mathbb{Z}_{2}$ spin liquid is not stable in the Heisenberg model on the breathing kagome lattice, and that its 
occurrence, as reported in a previous variational Monte Carlo study~\cite{Schaffer2017}, is a finite-size artifact. We emphasize that, in the 
isotropic case, the pairing terms are essentially vanishing for $L \geqslant 12$, as already reported in Refs.~\cite{Iqbal2011b,Iqbal2016}. 
Correspondingly, the thermodynamic extrapolation of $\Delta E$ is found to be vanishing for $J_{\triangledown}/J_{\vartriangle} \geqslant 0.3$ 
(within the error bar) and for $J_{\triangledown}/J_{\vartriangle}=0.1$ (within two error bars). In the latter case, the extrapolated result 
is tiny anyway, i.e., $\Delta E = 0.00002(1)$.

At this point, we would like to make a brief comment on the optimization procedure, which is particularly relevant for the isotropic point.
In particular, it has been suggested that finite pairing amplitudes are obtained up to large system sizes and in the thermodynamic 
limit~\cite{Li2016}, in contrast to what we have previously obtained~\cite{Iqbal2011b,Iqbal2016}. Indeed, on each size, it is possible to stabilize 
finite values of the pairing terms ($\zeta$ and $\Delta_{2}$), whenever the chemical potential $\mu$ does not correspond to the one of the Dirac 
state. However, once $\mu$ is correctly placed (i.e., within the highest occupied and the lowest unoccupied levels of the Dirac spectrum on each 
finite cluster), all the pairing amplitudes optimize to zero (within the error bar) for $L \geqslant 12$. In any case, also when the chemical 
potential is misplaced (and finite values of the pairings are obtained), the energy gain $\Delta E$ is still negligibly small on any finite system 
and scales to zero (within error bars) in the thermodynamic limit. Therefore, for understanding whether a gap opens up or not in reality, it is not 
sufficient to analyze the size scaling of the variational parameters alone, but rather a complete study of the energy gain on large finite systems 
together with a thermodynamic extrapolation must be afforded.
 
\begin{figure}
\includegraphics[width=\columnwidth]{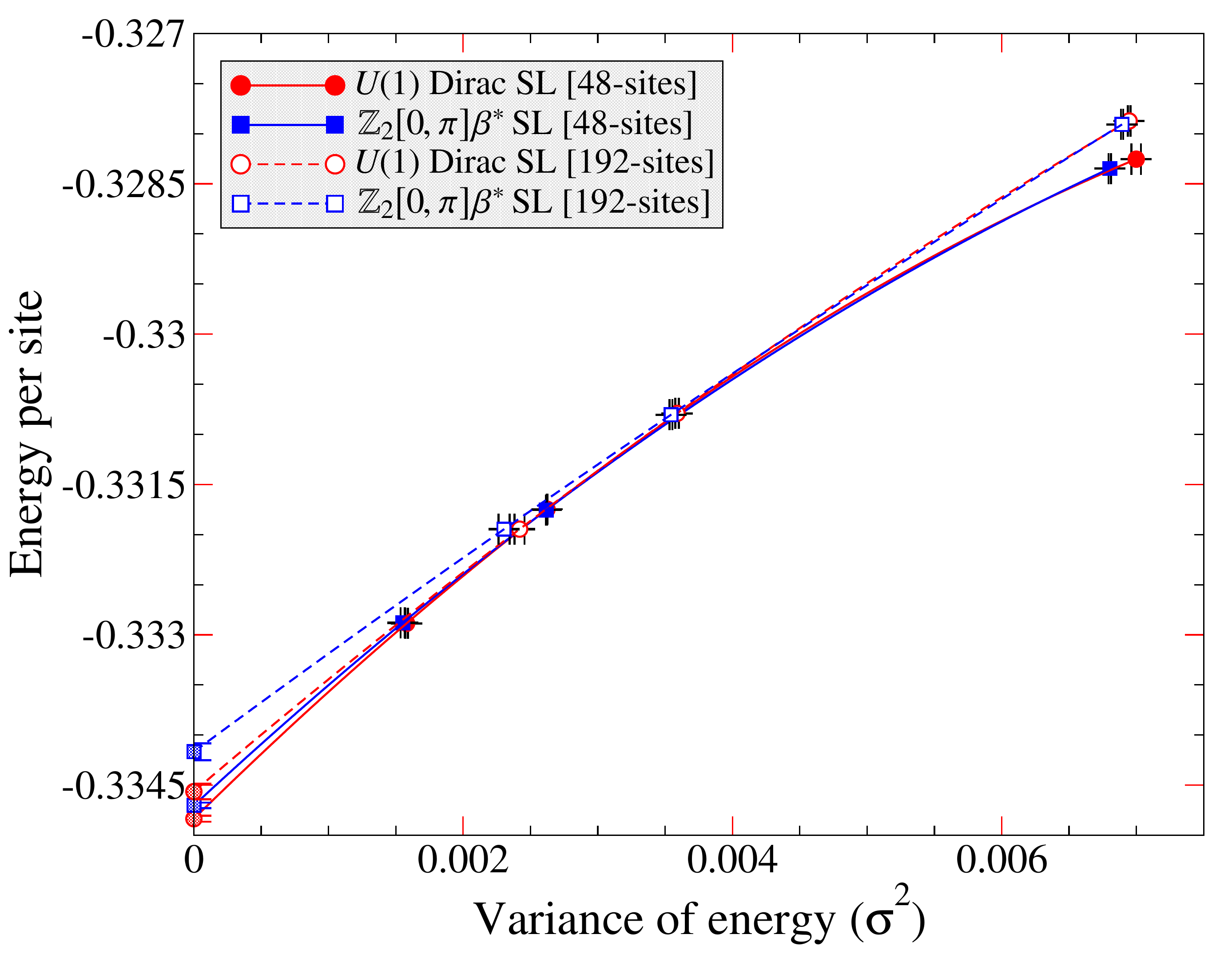}
\caption{\label{fig:LS}
For $J_{\triangledown}/J_{\vartriangle}=0.5$, the Lanczos step extrapolation (employing a quadratic fit) of the ground-state energy for the U(1)
Dirac and the $\mathbb{Z}_{2}[0,\pi]\beta^{*}$ states on the $48$- and $192$-site clusters.}
\end{figure}

The stability of the U(1) Dirac spin liquid with respect to the opening of a (topological) gap leading to the formation of a  $\mathbb{Z}_{2}$ 
state is  not an artifact of the variational approach. In order to prove this statement, we have performed one and two Lanczos steps on both the 
gapless U(1) and gapped $\mathbb{Z}_{2}$ states for $L=4$ and $8$ clusters at a given $J_{\triangledown}/J_{\vartriangle}=0.5$, also 
performing the zero-variance extrapolation that allows us to get a (nonvariational) estimation of the exact ground-state energy. The results 
are shown in Fig.~\ref{fig:LS} [see also the Supplemental Material (Ref.~\cite{supp})] and a few aspects should be stressed. First of all, we must emphasize that 
the finite-size energy gain of the $\mathbb{Z}_{2}$ \emph{Ansatz} decreases from $p=0$ to $p=2$, suggesting the fact that the fermionic pairing does not 
reflect the correct way to improve the original U(1) state. Moreover, the zero-variance extrapolated estimate of the energy for the U(1) Dirac 
state is slightly lower compared to the $\mathbb{Z}_{2}[0,\pi]\beta^{*}$ state on the $48$-site cluster, and this difference in energy increases 
on the $192$-site cluster, implying that the $\mathbb{Z}_{2}[0,\pi]\beta^{*}$ wave function performs worse with increasing system size. Even though 
an accurate extrapolation to the thermodynamic limit of the zero-variance energy is beyond the goals of the present work, we are confident that 
these results will be important for future comparisons that employ complementary numerical methods.

\subsection{Strong breathing anisotropy limit}

For completeness, we now focus on the strong anisotropy limit $J_{\triangledown}\ll J_{\vartriangle}$ where other states compete with the U(1) 
spin liquid. In particular, we shall investigate (i) the simplex topological RVB liquid (which can be written as a simple PEPS) and (ii) a VBC 
that is adiabatically connected to the projected U(1) state.  

\subsubsection{Competition with the simplex RVB liquid}

\begin{figure}
\includegraphics[width=\columnwidth]{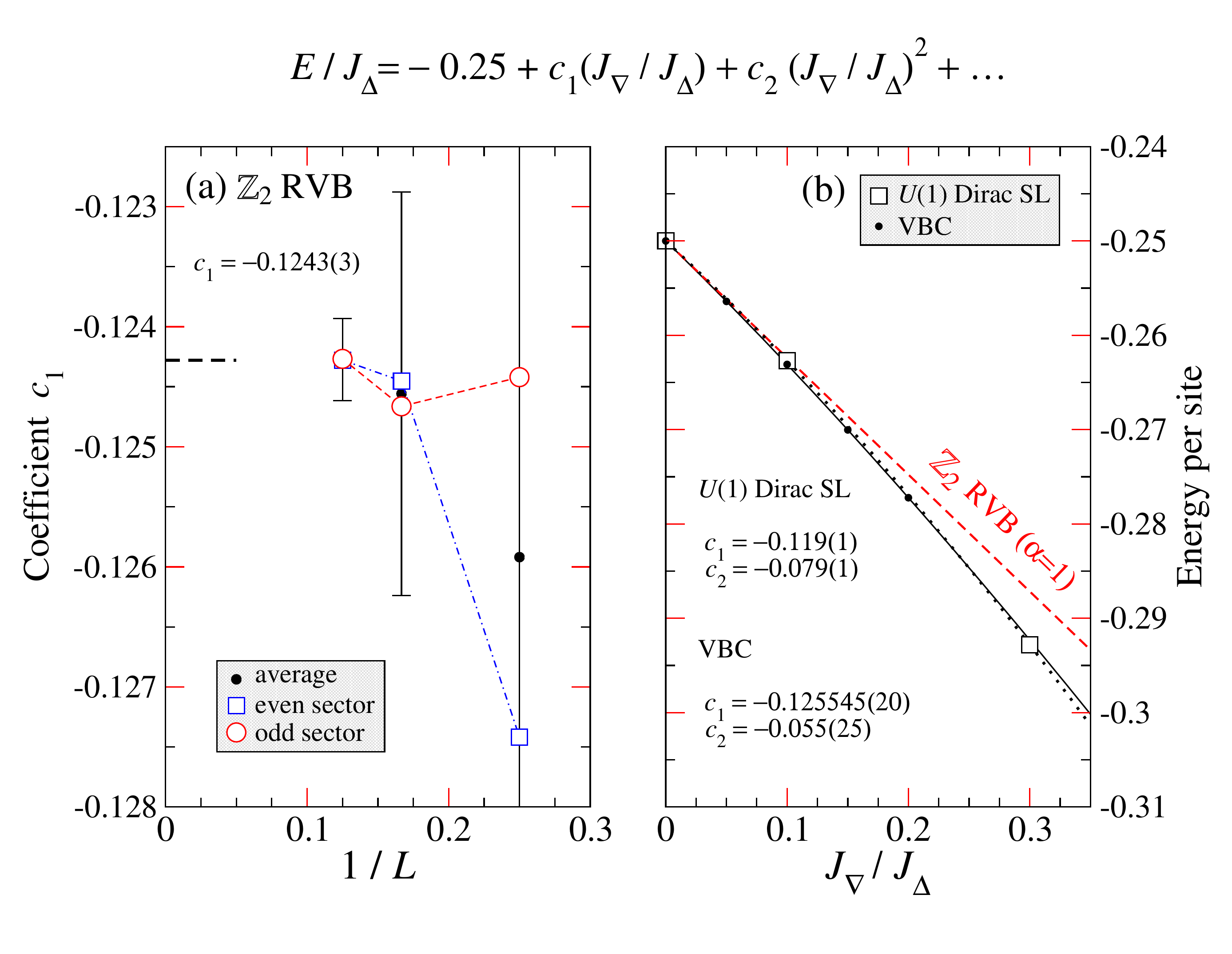}
\caption{\label{fig:Taylor}
(a) Coefficient $c_1$ (extracted from the energy at $\alpha=1$ of infinitely long cylinders of finite circumference $L$) plotted versus $1/L$. 
To minimize finite size effects, we consider the average over the two topological sectors. The error bars represent the energy difference (in 
units of $J_\triangledown$) between strong and weak bonds in the down triangles (nematicity), giving a tight bracketing of the extrapolation. 
(b) Energy (per site) in units of $J_{\vartriangle}$ vs $J_{\triangledown}/J_{\vartriangle}$ of (i) the U(1) wave function, (ii) the simplex 
RVB at fixed $\alpha=1$, and (iii) the optimal VBC state. Fits up to second and third order in $J_{\triangledown}/J_{\vartriangle}$ are used for 
the U(1) (dashed line) and the VBC (full line) states to extract the respective $c_1$ and $c_2$ parameters (with error bars). The VBC has lower 
energy up to $J_{\triangledown}/J_{\vartriangle} \approx 0.25$.}
\end{figure}

Here, we consider the simplex RVB written as a PEPS~\cite{Poilblanc2013a} and consider a Taylor expansion of the energy per site (in units of 
$J_{\vartriangle}$) in the strong anisotropy limit:
\begin{equation}\label{eq:taylor}
\frac{E}{J_{\vartriangle}}=-0.25+c_1 \frac{J_{\triangledown}}{J_{\vartriangle}} + c_2 \left (\frac{J_{\triangledown}}{J_{\vartriangle}} \right)^2 
+ \cdots \, .
\end{equation}
The constant $-0.25$ and the coefficient $c_1$ of the linear term are captured by setting $\alpha=1$ appearing in the operator 
$\mathbb{I}-\alpha \mathbb{P}_{3/2}$ acting on the up-pointing triangles (hence projecting exactly on the $S=1/2$ manifold of all up-pointing 
triangles). Note, however, that an optimization over the parameter $\alpha$ would be required at finite $J_{\triangledown}$ (and to get higher order 
terms in the Taylor expansion). From the energy per site $E=-0.25J_{\vartriangle}+c_1(L)J_{\triangledown}$ that is obtained for $\alpha=1$ on 
\emph{infinitely long} (vertical) cylinders of perimeter $L=4$, $6$, and $8$ unit cells (in each even or odd topological sector), we can extract the 
coefficient $c_1(L)$. Then, by performing the extrapolation $L\to\infty$ as shown in Fig.~\ref{fig:Taylor}(a), we obtain $c_1 \simeq -0.1243(3)$. 
Instead, a fit of the energy of the U(1) state gives $c_1\simeq -0.119(1)$, definitely above the value of the simplex RVB; see 
Fig.~\ref{fig:Taylor}(b). This implies that the simplex RVB has a lower energy than the U(1) wave function at a sufficiently small value of the 
coupling $J_{\triangledown}/J_{\vartriangle}$, whatever the respective values of the coefficient $c_2$ of the quadratic term. 

\begin{figure}
\includegraphics[width=\columnwidth]{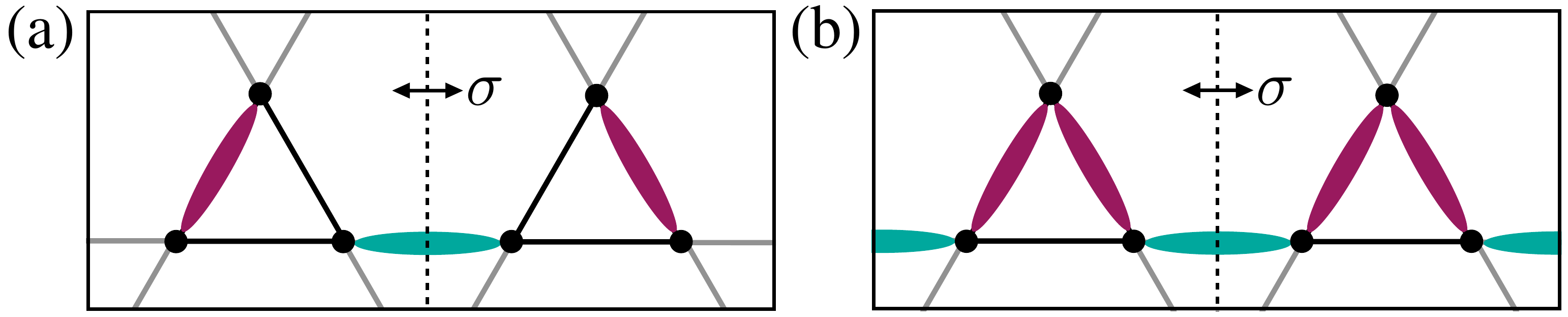}
\caption{\label{fig:VBC}
Schematic illustration of nearest-neighbor hopping amplitudes of the auxiliary Hamiltonian~[Eq.~(\ref{eq:MF-Z2})] for (a) VBC and (b) nematic 
states (no pairing terms are considered here): maroon (green) bonds within the up-pointing (down-pointing) triangles are stronger compared to 
the black (gray) bonds within the same triangles.}
\end{figure}

\begin{figure}[b]
\includegraphics[width=\columnwidth]{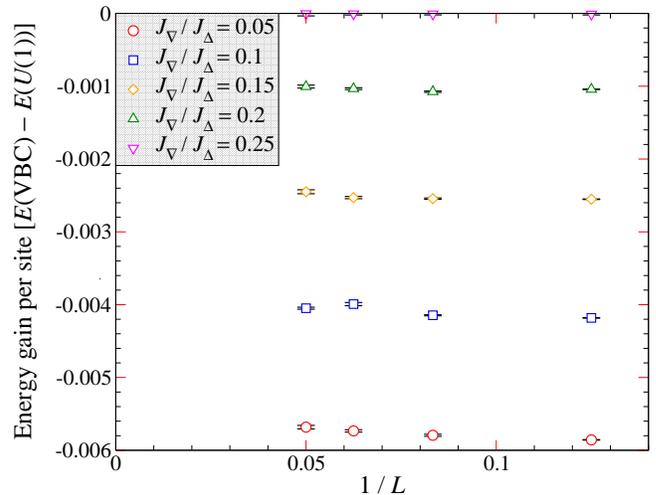}
\caption{\label{fig:VBC2}
For different values of the breathing anisotropy $J_{\triangledown}/J_{\vartriangle}$, the finite-size scaling of the energy gain (in units of 
$J_{\triangledown}$) of the six-site unit-cell VBC with respect to the U(1) Dirac spin liquid, i.e., $[E({\rm VBC}){-}E(\rm{U}(1))]/J_{\triangledown}$.
The clusters considered are $L=8$, $12$, $16$, and $20$.}
\end{figure}

\subsubsection{Evidence of a VBC ground state}

Now, we address the issue of the stability of the U(1) Dirac spin liquid towards dimerizing into a VBC. For simplicity, we choose a VBC with 
a unit cell of six-sites, i.e., composed of two geometrical unit cells, and impose a chosen pattern of amplitude modulation of nearest-neighbor 
hoppings on top of the uniform U(1) state; see Fig.~\ref{fig:VBC}(a). This dimer pattern breaks both the translational and the threefold rotational 
symmetry of the lattice, but preserves the reflection symmetry about an axis perpendicular to the primitive lattice vector {\bf a}$_1$.
Therefore, the VBC wave function has two different hopping amplitudes within up-pointing triangles, i.e., the maroon (strong) and black (weak) 
bonds, and also down-pointing triangles, i.e., green (strong) and gray (weak) bonds. This results in an enlarged variational parameter space and 
hence allows for potential lowering of energy. We optimize the VBC wave function for various values of the breathing anisotropy and find that, 
starting from the isotropic limit down to $J_{\triangledown}/J_{\vartriangle}\approx 0.25$, the optimization yields back the uniform U(1) spin 
liquid as the lowest energy state. Then, for $J_{\triangledown}/J_{\vartriangle}\lesssim 0.25$, the optimization of the VBC wave function yields 
an energy which is significantly lower compared to the U(1) Dirac state [see Fig.~\ref{fig:VBC2} and the Supplemental Material (Ref.~\cite{supp})]; 
therefore, the resulting wave function is characterized by a strong dimerization of the hopping amplitudes, with the maroon and green bonds [in 
Fig.~\ref{fig:VBC}(a)] being considerably stronger compared to the black and gray bonds. Most importantly, we find that the gain in the energy of 
the VBC with respect to the U(1) Dirac state, i.e., $E({\rm VBC})-E(\rm{U}(1))$, stays essentially constant with increasing system size from $L=8$ 
to $L=20$ (see Fig.~\ref{fig:VBC2}) pointing to the fact that the VBC wave function does not lose accuracy as $N \to \infty$, i.e., it is 
size consistent (unlike the gapped $\mathbb{Z}_{2}$ spin liquid). The variational energy of the optimal VBC state is also slightly lower
than the simplex RVB state that is constructed by using PEPS; see the analysis on the Taylor expansion of Eq.~(\ref{eq:taylor}) reported in 
Fig.~\ref{fig:Taylor}(b). These results thus provide strong evidence for a VBC ground state of the model in the regime 
$0<J_{\triangledown}/J_{\vartriangle}\lesssim 0.25$. 

We would like to mention that consideration of VBCs with larger unit cell with $12$ or $36$ sites, as defined in Refs.~\cite{Iqbal2012,Iqbal2011a}, 
and their optimization could possibly lead to further lowering of energy due to the enlargement of variational space; nonetheless, the fact that 
already for a six-site unit-cell VBC we obtain an appreciable and size-consistent energy gain is conclusive proof enough of a VBC ordered ground 
state in this parameter regime.

\subsubsection{Search for nematic order} 

We finally consider the case of a lattice-nematic state, which only breaks the threefold lattice rotational symmetry but preserves the
translational symmetry; see Fig.~\ref{fig:VBC}(b). By optimizing such a case for various values of the breathing anisotropy and starting from 
different points in variational parameter space (i.e., having different hopping amplitude modulations), we find that the optimization always 
returns back to the uniform U(1) Dirac state as the lowest energy one. In particular, in the regime of strong anisotropy, this points to the 
fact that in order to gain energy with respect to the U(1) spin liquid, it is crucial to break translational symmetry along with rotations. 
This fact is in contrast to the results obtained by the DMRG approach in Ref.~\cite{Repellin2017}, which claimed a pure lattice-nematic without 
any translational symmetry breaking. We want to stress that the simplex RVB wave function obtained within PEPS also showed nematicity [see 
Fig.~\ref{fig:Taylor}(a)]; however, this is an artifact induced by finite-perimeter cylinders (manifesting itself in the spatial anisotropy 
of spin-spin correlations) and drops off with increasing perimeter. In addition, there is no further energy gain by allowing a nematic bias
in the $R_{\vartriangle}$ tensor.
      
\section{Conclusions}\label{sec:concl}

We have investigated the nature of the ground state of the $S=1/2$ Heisenberg antiferromagnet on the breathing kagome lattice employing 
Gutzwiller projected wave functions analyzed with variational Monte Carlo methods. Based on high-accuracy and large-scale calculations, 
supplemented by a finite-size scaling analysis, we showed that the true thermodynamic ground state is a U(1) Dirac spin liquid for a wide span 
of breathing anisotropies, starting from (and including) the isotropic point $J_{\triangledown}/J_{\vartriangle}=1$ down to large anisotropies 
$J_{\triangledown}/J_{\vartriangle} \approx 0.25$. Our findings concerning the remarkable stability, robustness, and extent of the U(1) Dirac 
spin liquid are in excellent agreement with those from a recent DMRG study~\cite{Repellin2017}. The results are of direct relevance to the 
breathing kagome material vanadium oxyfluoride DQVOF, as the strength of breathing anisotropy estimated using series expansion is 
$J_{\triangledown}/J_{\vartriangle}=0.55(4)$~\cite{Orain2017}, which securely places DQVOF inside the regime of stability of the U(1) Dirac 
state. Our results are thus consistent with the gapless spin liquid behavior observed in spin-lattice ($T_{1}$) measurements~\cite{Orain2017} 
and lend support to the view that spin liquid behavior observed in DQVOF is likely to be intrinsic to the breathing kagome lattice. In addition,
our results would suggest that couplings between the $S=1/2$ V$^{4+}$ ions within the breathing kagome planes to the inter-layer $S=1$ V$^{3+}$ 
ions is not a necessary ingredient to generate spin liquid behavior.

In the regime of strong breathing anisotropy $J_{\triangledown}\ll J_{\vartriangle}$, we revealed the presence of a phase transition whereby the 
U(1) Dirac spin liquid undergoes a dimer instability and gives way to a VBC ground state for $J_{\triangledown}/J_{\vartriangle}\lesssim 0.25$. 
This finding is at variance with that from DMRG~\cite{Repellin2017}, which claimed a pure lattice-nematic state that preserves translations. 
Nonetheless, the remarkable agreement between the conclusions obtained from variational Monte Carlo and DMRG on the nature and extent of the 
ground state in a wide span of parameter space represents a milestone which hitherto could not be foreseen. It also highlights the quantitative 
and qualitative accuracy of projected fermionic wave functions (while only involving a few parameters) for spin models hosting a spin liquid 
ground state.  

\begin{acknowledgments} 
We thank M. Mambrini and F. Mila for helpful discussions. We acknowledge the kind hospitality and stimulating environment of the Centro de Ciencias de Benasque Pedro Pascual, Benasque, during the workshop ``Entanglement in Strongly Correlated Systems'' where this project was initiated. Y.I. and R.T. gratefully acknowledge the Gauss Centre for Supercomputing e.V. for funding this project by providing computing time on the GCS Supercomputer SuperMUC at Leibniz Supercomputing Centre (LRZ). D.P. acknowledges support from the French Research Council (ANR) under the NQPTP ANR-0406-01 grant and CALMIP (Toulouse) for CPU time on the EOS Supercomputer. R.T. acknowledges support through ERC-StG-TOPOLECTRICS-336012, DFG SFB 1170, and DFG SPP 1666.
\end{acknowledgments}

\newpage



\newcommand{\beginsupplement}{%
        \setcounter{table}{0}
        \renewcommand{\thetable}{S\arabic{table}}%
        \setcounter{figure}{0}
        \renewcommand{\thefigure}{S\arabic{figure}}%
        \setcounter{equation}{0}
        \renewcommand{\theequation}{S\arabic{equation}}%
        \setcounter{page}{1}
     }
 
 \bibliographystyle{apsrev4-1}

\begin{thebibliography}{62}%
\makeatletter
\providecommand \@ifxundefined [1]{%
 \@ifx{#1\undefined}
}%
\providecommand \@ifnum [1]{%
 \ifnum #1\expandafter \@firstoftwo
 \else \expandafter \@secondoftwo
 \fi
}%
\providecommand \@ifx [1]{%
 \ifx #1\expandafter \@firstoftwo
 \else \expandafter \@secondoftwo
 \fi
}%
\providecommand \natexlab [1]{#1}%
\providecommand \enquote  [1]{``#1''}%
\providecommand \bibnamefont  [1]{#1}%
\providecommand \bibfnamefont [1]{#1}%
\providecommand \citenamefont [1]{#1}%
\providecommand \href@noop [0]{\@secondoftwo}%
\providecommand \href [0]{\begingroup \@sanitize@url \@href}%
\providecommand \@href[1]{\@@startlink{#1}\@@href}%
\providecommand \@@href[1]{\endgroup#1\@@endlink}%
\providecommand \@sanitize@url [0]{\catcode `\\12\catcode `\$12\catcode
  `\&12\catcode `\#12\catcode `\^12\catcode `\_12\catcode `\%12\relax}%
\providecommand \@@startlink[1]{}%
\providecommand \@@endlink[0]{}%
\providecommand \url  [0]{\begingroup\@sanitize@url \@url }%
\providecommand \@url [1]{\endgroup\@href {#1}{\urlprefix }}%
\providecommand \urlprefix  [0]{URL }%
\providecommand \Eprint [0]{\href }%
\providecommand \doibase [0]{http://dx.doi.org/}%
\providecommand \selectlanguage [0]{\@gobble}%
\providecommand \bibinfo  [0]{\@secondoftwo}%
\providecommand \bibfield  [0]{\@secondoftwo}%
\providecommand \translation [1]{[#1]}%
\providecommand \BibitemOpen [0]{}%
\providecommand \bibitemStop [0]{}%
\providecommand \bibitemNoStop [0]{.\EOS\space}%
\providecommand \EOS [0]{\spacefactor3000\relax}%
\providecommand \BibitemShut  [1]{\csname bibitem#1\endcsname}%
\let\auto@bib@innerbib\@empty
\bibitem [{\citenamefont {Pomeranchuk}(1941)}]{Pomeranchuk1941}%
  \BibitemOpen
  \bibfield  {author} {\bibinfo {author} {\bibfnamefont {I.}~\bibnamefont
  {Pomeranchuk}},\ }\href@noop {} {\bibfield  {journal} {\bibinfo  {journal}
  {Zh. Eksp. Teor. Fiz.}\ }\textbf {\bibinfo {volume} {11}},\ \bibinfo {pages}
  {226} (\bibinfo {year} {1941})}\BibitemShut {NoStop}%
\bibitem [{\citenamefont {Balents}(2010)}]{Balents2010}%
  \BibitemOpen
  \bibfield  {author} {\bibinfo {author} {\bibfnamefont {L.}~\bibnamefont
  {Balents}},\ }\href {http://dx.doi.org/10.1038/nature08917} {\bibfield
  {journal} {\bibinfo  {journal} {Nature~(London)}\ }\textbf {\bibinfo {volume}
  {464}},\ \bibinfo {pages} {199} (\bibinfo {year} {2010})}\BibitemShut
  {NoStop}%
\bibitem [{\citenamefont {Savary}\ and\ \citenamefont
  {Balents}(2017)}]{Savary2017}%
  \BibitemOpen
  \bibfield  {author} {\bibinfo {author} {\bibfnamefont {L.}~\bibnamefont
  {Savary}}\ and\ \bibinfo {author} {\bibfnamefont {L.}~\bibnamefont
  {Balents}},\ }\href {http://stacks.iop.org/0034-4885/80/i=1/a=016502}
  {\bibfield  {journal} {\bibinfo  {journal} {Rep. Prog. Phys.}\ }\textbf
  {\bibinfo {volume} {80}},\ \bibinfo {pages} {016502} (\bibinfo {year}
  {2017})}\BibitemShut {NoStop}%
\bibitem [{\citenamefont {Zhou}\ \emph {et~al.}(2017)\citenamefont {Zhou},
  \citenamefont {Kanoda},\ and\ \citenamefont {Ng}}]{Zhou2017}%
  \BibitemOpen
  \bibfield  {author} {\bibinfo {author} {\bibfnamefont {Y.}~\bibnamefont
  {Zhou}}, \bibinfo {author} {\bibfnamefont {K.}~\bibnamefont {Kanoda}}, \ and\
  \bibinfo {author} {\bibfnamefont {T.-K.}\ \bibnamefont {Ng}},\ }\href
  {\doibase 10.1103/RevModPhys.89.025003} {\bibfield  {journal} {\bibinfo
  {journal} {Rev. Mod. Phys.}\ }\textbf {\bibinfo {volume} {89}},\ \bibinfo
  {pages} {025003} (\bibinfo {year} {2017})}\BibitemShut {NoStop}%
\bibitem [{\citenamefont {Zeng}\ and\ \citenamefont {Elser}(1990)}]{Zeng1990}%
  \BibitemOpen
  \bibfield  {author} {\bibinfo {author} {\bibfnamefont {C.}~\bibnamefont
  {Zeng}}\ and\ \bibinfo {author} {\bibfnamefont {V.}~\bibnamefont {Elser}},\
  }\href {\doibase 10.1103/PhysRevB.42.8436} {\bibfield  {journal} {\bibinfo
  {journal} {Phys. Rev. B}\ }\textbf {\bibinfo {volume} {42}},\ \bibinfo
  {pages} {8436} (\bibinfo {year} {1990})}\BibitemShut {NoStop}%
\bibitem [{\citenamefont {Sachdev}(1992)}]{Sachdev1992}%
  \BibitemOpen
  \bibfield  {author} {\bibinfo {author} {\bibfnamefont {S.}~\bibnamefont
  {Sachdev}},\ }\href {\doibase 10.1103/PhysRevB.45.12377} {\bibfield
  {journal} {\bibinfo  {journal} {Phys. Rev. B}\ }\textbf {\bibinfo {volume}
  {45}},\ \bibinfo {pages} {12377} (\bibinfo {year} {1992})}\BibitemShut
  {NoStop}%
\bibitem [{\citenamefont {Lecheminant}\ \emph {et~al.}(1997)\citenamefont
  {Lecheminant}, \citenamefont {Bernu}, \citenamefont {Lhuillier},
  \citenamefont {Pierre},\ and\ \citenamefont {Sindzingre}}]{Lecheminant1997}%
  \BibitemOpen
  \bibfield  {author} {\bibinfo {author} {\bibfnamefont {P.}~\bibnamefont
  {Lecheminant}}, \bibinfo {author} {\bibfnamefont {B.}~\bibnamefont {Bernu}},
  \bibinfo {author} {\bibfnamefont {C.}~\bibnamefont {Lhuillier}}, \bibinfo
  {author} {\bibfnamefont {L.}~\bibnamefont {Pierre}}, \ and\ \bibinfo {author}
  {\bibfnamefont {P.}~\bibnamefont {Sindzingre}},\ }\href {\doibase
  10.1103/PhysRevB.56.2521} {\bibfield  {journal} {\bibinfo  {journal} {Phys.
  Rev. B}\ }\textbf {\bibinfo {volume} {56}},\ \bibinfo {pages} {2521}
  (\bibinfo {year} {1997})}\BibitemShut {NoStop}%
\bibitem [{\citenamefont {Shores}\ \emph {et~al.}(2005)\citenamefont {Shores},
  \citenamefont {Nytko}, \citenamefont {Bartlett},\ and\ \citenamefont
  {Nocera}}]{Shores2005}%
  \BibitemOpen
  \bibfield  {author} {\bibinfo {author} {\bibfnamefont {M.~P.}\ \bibnamefont
  {Shores}}, \bibinfo {author} {\bibfnamefont {E.~A.}\ \bibnamefont {Nytko}},
  \bibinfo {author} {\bibfnamefont {B.~M.}\ \bibnamefont {Bartlett}}, \ and\
  \bibinfo {author} {\bibfnamefont {D.~G.}\ \bibnamefont {Nocera}},\ }\href
  {\doibase 10.1021/ja053891p} {\bibfield  {journal} {\bibinfo  {journal} {J.
  Am. Chem. Soc.}\ }\textbf {\bibinfo {volume} {127}},\ \bibinfo {pages}
  {13462} (\bibinfo {year} {2005})}\BibitemShut {NoStop}%
\bibitem [{\citenamefont {Mendels}\ \emph {et~al.}(2007)\citenamefont
  {Mendels}, \citenamefont {Bert}, \citenamefont {de~Vries}, \citenamefont
  {Olariu}, \citenamefont {Harrison}, \citenamefont {Duc}, \citenamefont
  {Trombe}, \citenamefont {Lord}, \citenamefont {Amato},\ and\ \citenamefont
  {Baines}}]{Mendels2007}%
  \BibitemOpen
  \bibfield  {author} {\bibinfo {author} {\bibfnamefont {P.}~\bibnamefont
  {Mendels}}, \bibinfo {author} {\bibfnamefont {F.}~\bibnamefont {Bert}},
  \bibinfo {author} {\bibfnamefont {M.~A.}\ \bibnamefont {de~Vries}}, \bibinfo
  {author} {\bibfnamefont {A.}~\bibnamefont {Olariu}}, \bibinfo {author}
  {\bibfnamefont {A.}~\bibnamefont {Harrison}}, \bibinfo {author}
  {\bibfnamefont {F.}~\bibnamefont {Duc}}, \bibinfo {author} {\bibfnamefont
  {J.~C.}\ \bibnamefont {Trombe}}, \bibinfo {author} {\bibfnamefont {J.~S.}\
  \bibnamefont {Lord}}, \bibinfo {author} {\bibfnamefont {A.}~\bibnamefont
  {Amato}}, \ and\ \bibinfo {author} {\bibfnamefont {C.}~\bibnamefont
  {Baines}},\ }\href {\doibase 10.1103/PhysRevLett.98.077204} {\bibfield
  {journal} {\bibinfo  {journal} {Phys. Rev. Lett.}\ }\textbf {\bibinfo
  {volume} {98}},\ \bibinfo {pages} {077204} (\bibinfo {year}
  {2007})}\BibitemShut {NoStop}%
\bibitem [{\citenamefont {Helton}\ \emph {et~al.}(2007)\citenamefont {Helton},
  \citenamefont {Matan}, \citenamefont {Shores}, \citenamefont {Nytko},
  \citenamefont {Bartlett}, \citenamefont {Yoshida}, \citenamefont {Takano},
  \citenamefont {Suslov}, \citenamefont {Qiu}, \citenamefont {Chung},
  \citenamefont {Nocera},\ and\ \citenamefont {Lee}}]{Helton2007}%
  \BibitemOpen
  \bibfield  {author} {\bibinfo {author} {\bibfnamefont {J.~S.}\ \bibnamefont
  {Helton}}, \bibinfo {author} {\bibfnamefont {K.}~\bibnamefont {Matan}},
  \bibinfo {author} {\bibfnamefont {M.~P.}\ \bibnamefont {Shores}}, \bibinfo
  {author} {\bibfnamefont {E.~A.}\ \bibnamefont {Nytko}}, \bibinfo {author}
  {\bibfnamefont {B.~M.}\ \bibnamefont {Bartlett}}, \bibinfo {author}
  {\bibfnamefont {Y.}~\bibnamefont {Yoshida}}, \bibinfo {author} {\bibfnamefont
  {Y.}~\bibnamefont {Takano}}, \bibinfo {author} {\bibfnamefont
  {A.}~\bibnamefont {Suslov}}, \bibinfo {author} {\bibfnamefont
  {Y.}~\bibnamefont {Qiu}}, \bibinfo {author} {\bibfnamefont {J.-H.}\
  \bibnamefont {Chung}}, \bibinfo {author} {\bibfnamefont {D.~G.}\ \bibnamefont
  {Nocera}}, \ and\ \bibinfo {author} {\bibfnamefont {Y.~S.}\ \bibnamefont
  {Lee}},\ }\href {\doibase 10.1103/PhysRevLett.98.107204} {\bibfield
  {journal} {\bibinfo  {journal} {Phys. Rev. Lett.}\ }\textbf {\bibinfo
  {volume} {98}},\ \bibinfo {pages} {107204} (\bibinfo {year}
  {2007})}\BibitemShut {NoStop}%
\bibitem [{\citenamefont {Suttner}\ \emph {et~al.}(2014)\citenamefont
  {Suttner}, \citenamefont {Platt}, \citenamefont {Reuther},\ and\
  \citenamefont {Thomale}}]{Suttner2014}%
  \BibitemOpen
  \bibfield  {author} {\bibinfo {author} {\bibfnamefont {R.}~\bibnamefont
  {Suttner}}, \bibinfo {author} {\bibfnamefont {C.}~\bibnamefont {Platt}},
  \bibinfo {author} {\bibfnamefont {J.}~\bibnamefont {Reuther}}, \ and\
  \bibinfo {author} {\bibfnamefont {R.}~\bibnamefont {Thomale}},\ }\href
  {\doibase 10.1103/PhysRevB.89.020408} {\bibfield  {journal} {\bibinfo
  {journal} {Phys. Rev. B}\ }\textbf {\bibinfo {volume} {89}},\ \bibinfo
  {pages} {020408} (\bibinfo {year} {2014})}\BibitemShut {NoStop}%
\bibitem [{\citenamefont {Olariu}\ \emph {et~al.}(2008)\citenamefont {Olariu},
  \citenamefont {Mendels}, \citenamefont {Bert}, \citenamefont {Duc},
  \citenamefont {Trombe}, \citenamefont {de~Vries},\ and\ \citenamefont
  {Harrison}}]{Olariu2008}%
  \BibitemOpen
  \bibfield  {author} {\bibinfo {author} {\bibfnamefont {A.}~\bibnamefont
  {Olariu}}, \bibinfo {author} {\bibfnamefont {P.}~\bibnamefont {Mendels}},
  \bibinfo {author} {\bibfnamefont {F.}~\bibnamefont {Bert}}, \bibinfo {author}
  {\bibfnamefont {F.}~\bibnamefont {Duc}}, \bibinfo {author} {\bibfnamefont
  {J.~C.}\ \bibnamefont {Trombe}}, \bibinfo {author} {\bibfnamefont {M.~A.}\
  \bibnamefont {de~Vries}}, \ and\ \bibinfo {author} {\bibfnamefont
  {A.}~\bibnamefont {Harrison}},\ }\href {\doibase
  10.1103/PhysRevLett.100.087202} {\bibfield  {journal} {\bibinfo  {journal}
  {Phys. Rev. Lett.}\ }\textbf {\bibinfo {volume} {100}},\ \bibinfo {pages}
  {087202} (\bibinfo {year} {2008})}\BibitemShut {NoStop}%
\bibitem [{\citenamefont {Han}\ \emph {et~al.}(2012)\citenamefont {Han},
  \citenamefont {Helton}, \citenamefont {Chu}, \citenamefont {Nocera},
  \citenamefont {Rodriguez-Rivera}, \citenamefont {Broholm},\ and\
  \citenamefont {Lee}}]{Han2012}%
  \BibitemOpen
  \bibfield  {author} {\bibinfo {author} {\bibfnamefont {T.-H.}\ \bibnamefont
  {Han}}, \bibinfo {author} {\bibfnamefont {J.~S.}\ \bibnamefont {Helton}},
  \bibinfo {author} {\bibfnamefont {S.}~\bibnamefont {Chu}}, \bibinfo {author}
  {\bibfnamefont {D.~G.}\ \bibnamefont {Nocera}}, \bibinfo {author}
  {\bibfnamefont {J.~A.}\ \bibnamefont {Rodriguez-Rivera}}, \bibinfo {author}
  {\bibfnamefont {C.}~\bibnamefont {Broholm}}, \ and\ \bibinfo {author}
  {\bibfnamefont {Y.~S.}\ \bibnamefont {Lee}},\ }\href
  {http://dx.doi.org/10.1038/nature11659} {\bibfield  {journal} {\bibinfo
  {journal} {Nature (London)}\ }\textbf {\bibinfo {volume} {492}},\ \bibinfo
  {pages} {406} (\bibinfo {year} {2012})}\BibitemShut {NoStop}%
\bibitem [{\citenamefont {Fu}\ \emph {et~al.}(2015)\citenamefont {Fu},
  \citenamefont {Imai}, \citenamefont {Han},\ and\ \citenamefont
  {Lee}}]{Fu2015}%
  \BibitemOpen
  \bibfield  {author} {\bibinfo {author} {\bibfnamefont {M.}~\bibnamefont
  {Fu}}, \bibinfo {author} {\bibfnamefont {T.}~\bibnamefont {Imai}}, \bibinfo
  {author} {\bibfnamefont {T.-H.}\ \bibnamefont {Han}}, \ and\ \bibinfo
  {author} {\bibfnamefont {Y.~S.}\ \bibnamefont {Lee}},\ }\href {\doibase
  10.1126/science.aab2120} {\bibfield  {journal} {\bibinfo  {journal}
  {Science}\ }\textbf {\bibinfo {volume} {350}},\ \bibinfo {pages} {655}
  (\bibinfo {year} {2015})}\BibitemShut {NoStop}%
\bibitem [{\citenamefont {Jiang}\ \emph {et~al.}(2008)\citenamefont {Jiang},
  \citenamefont {Weng},\ and\ \citenamefont {Sheng}}]{Jiang2008}%
  \BibitemOpen
  \bibfield  {author} {\bibinfo {author} {\bibfnamefont {H.~C.}\ \bibnamefont
  {Jiang}}, \bibinfo {author} {\bibfnamefont {Z.~Y.}\ \bibnamefont {Weng}}, \
  and\ \bibinfo {author} {\bibfnamefont {D.~N.}\ \bibnamefont {Sheng}},\ }\href
  {\doibase 10.1103/PhysRevLett.101.117203} {\bibfield  {journal} {\bibinfo
  {journal} {Phys. Rev. Lett.}\ }\textbf {\bibinfo {volume} {101}},\ \bibinfo
  {pages} {117203} (\bibinfo {year} {2008})}\BibitemShut {NoStop}%
\bibitem [{\citenamefont {Yan}\ \emph {et~al.}(2011)\citenamefont {Yan},
  \citenamefont {Huse},\ and\ \citenamefont {White}}]{Yan2011}%
  \BibitemOpen
  \bibfield  {author} {\bibinfo {author} {\bibfnamefont {S.}~\bibnamefont
  {Yan}}, \bibinfo {author} {\bibfnamefont {D.~A.}\ \bibnamefont {Huse}}, \
  and\ \bibinfo {author} {\bibfnamefont {S.~R.}\ \bibnamefont {White}},\ }\href
  {\doibase 10.1126/science.1201080} {\bibfield  {journal} {\bibinfo  {journal}
  {Science}\ }\textbf {\bibinfo {volume} {332}},\ \bibinfo {pages} {1173}
  (\bibinfo {year} {2011})}\BibitemShut {NoStop}%
\bibitem [{\citenamefont {Depenbrock}\ \emph {et~al.}(2012)\citenamefont
  {Depenbrock}, \citenamefont {McCulloch},\ and\ \citenamefont
  {Schollw\"ock}}]{Depenbrock2012}%
  \BibitemOpen
  \bibfield  {author} {\bibinfo {author} {\bibfnamefont {S.}~\bibnamefont
  {Depenbrock}}, \bibinfo {author} {\bibfnamefont {I.~P.}\ \bibnamefont
  {McCulloch}}, \ and\ \bibinfo {author} {\bibfnamefont {U.}~\bibnamefont
  {Schollw\"ock}},\ }\href {\doibase 10.1103/PhysRevLett.109.067201} {\bibfield
   {journal} {\bibinfo  {journal} {Phys. Rev. Lett.}\ }\textbf {\bibinfo
  {volume} {109}},\ \bibinfo {pages} {067201} (\bibinfo {year}
  {2012})}\BibitemShut {NoStop}%
\bibitem [{\citenamefont {Ran}\ \emph {et~al.}(2007)\citenamefont {Ran},
  \citenamefont {Hermele}, \citenamefont {Lee},\ and\ \citenamefont
  {Wen}}]{Ran2007}%
  \BibitemOpen
  \bibfield  {author} {\bibinfo {author} {\bibfnamefont {Y.}~\bibnamefont
  {Ran}}, \bibinfo {author} {\bibfnamefont {M.}~\bibnamefont {Hermele}},
  \bibinfo {author} {\bibfnamefont {P.~A.}\ \bibnamefont {Lee}}, \ and\
  \bibinfo {author} {\bibfnamefont {X.-G.}\ \bibnamefont {Wen}},\ }\href
  {\doibase 10.1103/PhysRevLett.98.117205} {\bibfield  {journal} {\bibinfo
  {journal} {Phys. Rev. Lett.}\ }\textbf {\bibinfo {volume} {98}},\ \bibinfo
  {pages} {117205} (\bibinfo {year} {2007})}\BibitemShut {NoStop}%
\bibitem [{\citenamefont {Iqbal}\ \emph {et~al.}(2013)\citenamefont {Iqbal},
  \citenamefont {Becca}, \citenamefont {Sorella},\ and\ \citenamefont
  {Poilblanc}}]{Iqbal2013}%
  \BibitemOpen
  \bibfield  {author} {\bibinfo {author} {\bibfnamefont {Y.}~\bibnamefont
  {Iqbal}}, \bibinfo {author} {\bibfnamefont {F.}~\bibnamefont {Becca}},
  \bibinfo {author} {\bibfnamefont {S.}~\bibnamefont {Sorella}}, \ and\
  \bibinfo {author} {\bibfnamefont {D.}~\bibnamefont {Poilblanc}},\ }\href
  {\doibase 10.1103/PhysRevB.87.060405} {\bibfield  {journal} {\bibinfo
  {journal} {Phys. Rev. B}\ }\textbf {\bibinfo {volume} {87}},\ \bibinfo
  {pages} {060405} (\bibinfo {year} {2013})}\BibitemShut {NoStop}%
\bibitem [{\citenamefont {Iqbal}\ \emph {et~al.}(2014)\citenamefont {Iqbal},
  \citenamefont {Poilblanc},\ and\ \citenamefont {Becca}}]{Iqbal2014}%
  \BibitemOpen
  \bibfield  {author} {\bibinfo {author} {\bibfnamefont {Y.}~\bibnamefont
  {Iqbal}}, \bibinfo {author} {\bibfnamefont {D.}~\bibnamefont {Poilblanc}}, \
  and\ \bibinfo {author} {\bibfnamefont {F.}~\bibnamefont {Becca}},\ }\href
  {\doibase 10.1103/PhysRevB.89.020407} {\bibfield  {journal} {\bibinfo
  {journal} {Phys. Rev. B}\ }\textbf {\bibinfo {volume} {89}},\ \bibinfo
  {pages} {020407} (\bibinfo {year} {2014})}\BibitemShut {NoStop}%
\bibitem [{\citenamefont {Iqbal}\ \emph {et~al.}(2015)\citenamefont {Iqbal},
  \citenamefont {Poilblanc},\ and\ \citenamefont {Becca}}]{Iqbal2015}%
  \BibitemOpen
  \bibfield  {author} {\bibinfo {author} {\bibfnamefont {Y.}~\bibnamefont
  {Iqbal}}, \bibinfo {author} {\bibfnamefont {D.}~\bibnamefont {Poilblanc}}, \
  and\ \bibinfo {author} {\bibfnamefont {F.}~\bibnamefont {Becca}},\ }\href
  {\doibase 10.1103/PhysRevB.91.020402} {\bibfield  {journal} {\bibinfo
  {journal} {Phys. Rev. B}\ }\textbf {\bibinfo {volume} {91}},\ \bibinfo
  {pages} {020402} (\bibinfo {year} {2015})}\BibitemShut {NoStop}%
\bibitem [{\citenamefont {He}\ \emph {et~al.}(2017)\citenamefont {He},
  \citenamefont {Zaletel}, \citenamefont {Oshikawa},\ and\ \citenamefont
  {Pollmann}}]{He2017}%
  \BibitemOpen
  \bibfield  {author} {\bibinfo {author} {\bibfnamefont {Y.-C.}\ \bibnamefont
  {He}}, \bibinfo {author} {\bibfnamefont {M.~P.}\ \bibnamefont {Zaletel}},
  \bibinfo {author} {\bibfnamefont {M.}~\bibnamefont {Oshikawa}}, \ and\
  \bibinfo {author} {\bibfnamefont {F.}~\bibnamefont {Pollmann}},\ }\href
  {\doibase 10.1103/PhysRevX.7.031020} {\bibfield  {journal} {\bibinfo
  {journal} {Phys. Rev. X}\ }\textbf {\bibinfo {volume} {7}},\ \bibinfo {pages}
  {031020} (\bibinfo {year} {2017})}\BibitemShut {NoStop}%
\bibitem [{\citenamefont {{Zhu}}\ \emph {et~al.}(2018)\citenamefont {{Zhu}},
  \citenamefont {{Chen}}, \citenamefont {{He}},\ and\ \citenamefont
  {{Witczak-Krempa}}}]{Zhu-2018}%
  \BibitemOpen
  \bibfield  {author} {\bibinfo {author} {\bibfnamefont {W.}~\bibnamefont
  {{Zhu}}}, \bibinfo {author} {\bibfnamefont {X.}~\bibnamefont {{Chen}}},
  \bibinfo {author} {\bibfnamefont {Y.-C.}\ \bibnamefont {{He}}}, \ and\
  \bibinfo {author} {\bibfnamefont {W.}~\bibnamefont {{Witczak-Krempa}}},\
  }\href@noop {} {\bibfield  {journal} {\bibinfo  {journal} {ArXiv e-prints}\ }
  (\bibinfo {year} {2018})},\ \Eprint {http://arxiv.org/abs/1801.06177}
  {arXiv:1801.06177 [cond-mat.str-el]} \BibitemShut {NoStop}%
\bibitem [{\citenamefont {Liao}\ \emph {et~al.}(2017)\citenamefont {Liao},
  \citenamefont {Xie}, \citenamefont {Chen}, \citenamefont {Liu}, \citenamefont
  {Xie}, \citenamefont {Huang}, \citenamefont {Normand},\ and\ \citenamefont
  {Xiang}}]{Liao2017}%
  \BibitemOpen
  \bibfield  {author} {\bibinfo {author} {\bibfnamefont {H.~J.}\ \bibnamefont
  {Liao}}, \bibinfo {author} {\bibfnamefont {Z.~Y.}\ \bibnamefont {Xie}},
  \bibinfo {author} {\bibfnamefont {J.}~\bibnamefont {Chen}}, \bibinfo {author}
  {\bibfnamefont {Z.~Y.}\ \bibnamefont {Liu}}, \bibinfo {author} {\bibfnamefont
  {H.~D.}\ \bibnamefont {Xie}}, \bibinfo {author} {\bibfnamefont {R.~Z.}\
  \bibnamefont {Huang}}, \bibinfo {author} {\bibfnamefont {B.}~\bibnamefont
  {Normand}}, \ and\ \bibinfo {author} {\bibfnamefont {T.}~\bibnamefont
  {Xiang}},\ }\href {\doibase 10.1103/PhysRevLett.118.137202} {\bibfield
  {journal} {\bibinfo  {journal} {Phys. Rev. Lett.}\ }\textbf {\bibinfo
  {volume} {118}},\ \bibinfo {pages} {137202} (\bibinfo {year}
  {2017})}\BibitemShut {NoStop}%
\bibitem [{\citenamefont {Bert}\ \emph {et~al.}(2005)\citenamefont {Bert},
  \citenamefont {Bono}, \citenamefont {Mendels}, \citenamefont {Ladieu},
  \citenamefont {Duc}, \citenamefont {Trombe},\ and\ \citenamefont
  {Millet}}]{Bert2005}%
  \BibitemOpen
  \bibfield  {author} {\bibinfo {author} {\bibfnamefont {F.}~\bibnamefont
  {Bert}}, \bibinfo {author} {\bibfnamefont {D.}~\bibnamefont {Bono}}, \bibinfo
  {author} {\bibfnamefont {P.}~\bibnamefont {Mendels}}, \bibinfo {author}
  {\bibfnamefont {F.}~\bibnamefont {Ladieu}}, \bibinfo {author} {\bibfnamefont
  {F.}~\bibnamefont {Duc}}, \bibinfo {author} {\bibfnamefont {J.-C.}\
  \bibnamefont {Trombe}}, \ and\ \bibinfo {author} {\bibfnamefont
  {P.}~\bibnamefont {Millet}},\ }\href {\doibase 10.1103/PhysRevLett.95.087203}
  {\bibfield  {journal} {\bibinfo  {journal} {Phys. Rev. Lett.}\ }\textbf
  {\bibinfo {volume} {95}},\ \bibinfo {pages} {087203} (\bibinfo {year}
  {2005})}\BibitemShut {NoStop}%
\bibitem [{\citenamefont {Yavors'kii}\ \emph {et~al.}(2007)\citenamefont
  {Yavors'kii}, \citenamefont {Apel},\ and\ \citenamefont
  {Everts}}]{Yavorskii2007}%
  \BibitemOpen
  \bibfield  {author} {\bibinfo {author} {\bibfnamefont {T.}~\bibnamefont
  {Yavors'kii}}, \bibinfo {author} {\bibfnamefont {W.}~\bibnamefont {Apel}}, \
  and\ \bibinfo {author} {\bibfnamefont {H.-U.}\ \bibnamefont {Everts}},\
  }\href {\doibase 10.1103/PhysRevB.76.064430} {\bibfield  {journal} {\bibinfo
  {journal} {Phys. Rev. B}\ }\textbf {\bibinfo {volume} {76}},\ \bibinfo
  {pages} {064430} (\bibinfo {year} {2007})}\BibitemShut {NoStop}%
\bibitem [{\citenamefont {Wang}\ \emph {et~al.}(2007)\citenamefont {Wang},
  \citenamefont {Vishwanath},\ and\ \citenamefont {Kim}}]{Wang2007}%
  \BibitemOpen
  \bibfield  {author} {\bibinfo {author} {\bibfnamefont {F.}~\bibnamefont
  {Wang}}, \bibinfo {author} {\bibfnamefont {A.}~\bibnamefont {Vishwanath}}, \
  and\ \bibinfo {author} {\bibfnamefont {Y.~B.}\ \bibnamefont {Kim}},\ }\href
  {\doibase 10.1103/PhysRevB.76.094421} {\bibfield  {journal} {\bibinfo
  {journal} {Phys. Rev. B}\ }\textbf {\bibinfo {volume} {76}},\ \bibinfo
  {pages} {094421} (\bibinfo {year} {2007})}\BibitemShut {NoStop}%
\bibitem [{\citenamefont {Yoshida}\ \emph {et~al.}(2009)\citenamefont
  {Yoshida}, \citenamefont {Takigawa}, \citenamefont {Yoshida}, \citenamefont
  {Okamoto},\ and\ \citenamefont {Hiroi}}]{Yoshida2009}%
  \BibitemOpen
  \bibfield  {author} {\bibinfo {author} {\bibfnamefont {M.}~\bibnamefont
  {Yoshida}}, \bibinfo {author} {\bibfnamefont {M.}~\bibnamefont {Takigawa}},
  \bibinfo {author} {\bibfnamefont {H.}~\bibnamefont {Yoshida}}, \bibinfo
  {author} {\bibfnamefont {Y.}~\bibnamefont {Okamoto}}, \ and\ \bibinfo
  {author} {\bibfnamefont {Z.}~\bibnamefont {Hiroi}},\ }\href {\doibase
  10.1103/PhysRevLett.103.077207} {\bibfield  {journal} {\bibinfo  {journal}
  {Phys. Rev. Lett.}\ }\textbf {\bibinfo {volume} {103}},\ \bibinfo {pages}
  {077207} (\bibinfo {year} {2009})}\BibitemShut {NoStop}%
\bibitem [{\citenamefont {Janson}\ \emph {et~al.}(2010)\citenamefont {Janson},
  \citenamefont {Richter}, \citenamefont {Sindzingre},\ and\ \citenamefont
  {Rosner}}]{Janson2010}%
  \BibitemOpen
  \bibfield  {author} {\bibinfo {author} {\bibfnamefont {O.}~\bibnamefont
  {Janson}}, \bibinfo {author} {\bibfnamefont {J.}~\bibnamefont {Richter}},
  \bibinfo {author} {\bibfnamefont {P.}~\bibnamefont {Sindzingre}}, \ and\
  \bibinfo {author} {\bibfnamefont {H.}~\bibnamefont {Rosner}},\ }\href
  {\doibase 10.1103/PhysRevB.82.104434} {\bibfield  {journal} {\bibinfo
  {journal} {Phys. Rev. B}\ }\textbf {\bibinfo {volume} {82}},\ \bibinfo
  {pages} {104434} (\bibinfo {year} {2010})}\BibitemShut {NoStop}%
\bibitem [{\citenamefont {Nilsen}\ \emph {et~al.}(2011)\citenamefont {Nilsen},
  \citenamefont {Coomer}, \citenamefont {de~Vries}, \citenamefont {Stewart},
  \citenamefont {Deen}, \citenamefont {Harrison},\ and\ \citenamefont
  {R\o{}nnow}}]{Nilsen2011}%
  \BibitemOpen
  \bibfield  {author} {\bibinfo {author} {\bibfnamefont {G.~J.}\ \bibnamefont
  {Nilsen}}, \bibinfo {author} {\bibfnamefont {F.~C.}\ \bibnamefont {Coomer}},
  \bibinfo {author} {\bibfnamefont {M.~A.}\ \bibnamefont {de~Vries}}, \bibinfo
  {author} {\bibfnamefont {J.~R.}\ \bibnamefont {Stewart}}, \bibinfo {author}
  {\bibfnamefont {P.~P.}\ \bibnamefont {Deen}}, \bibinfo {author}
  {\bibfnamefont {A.}~\bibnamefont {Harrison}}, \ and\ \bibinfo {author}
  {\bibfnamefont {H.~M.}\ \bibnamefont {R\o{}nnow}},\ }\href {\doibase
  10.1103/PhysRevB.84.172401} {\bibfield  {journal} {\bibinfo  {journal} {Phys.
  Rev. B}\ }\textbf {\bibinfo {volume} {84}},\ \bibinfo {pages} {172401}
  (\bibinfo {year} {2011})}\BibitemShut {NoStop}%
\bibitem [{\citenamefont {Janson}\ \emph {et~al.}(2016)\citenamefont {Janson},
  \citenamefont {Furukawa}, \citenamefont {Momoi}, \citenamefont {Sindzingre},
  \citenamefont {Richter},\ and\ \citenamefont {Held}}]{Janson2016}%
  \BibitemOpen
  \bibfield  {author} {\bibinfo {author} {\bibfnamefont {O.}~\bibnamefont
  {Janson}}, \bibinfo {author} {\bibfnamefont {S.}~\bibnamefont {Furukawa}},
  \bibinfo {author} {\bibfnamefont {T.}~\bibnamefont {Momoi}}, \bibinfo
  {author} {\bibfnamefont {P.}~\bibnamefont {Sindzingre}}, \bibinfo {author}
  {\bibfnamefont {J.}~\bibnamefont {Richter}}, \ and\ \bibinfo {author}
  {\bibfnamefont {K.}~\bibnamefont {Held}},\ }\href {\doibase
  10.1103/PhysRevLett.117.037206} {\bibfield  {journal} {\bibinfo  {journal}
  {Phys. Rev. Lett.}\ }\textbf {\bibinfo {volume} {117}},\ \bibinfo {pages}
  {037206} (\bibinfo {year} {2016})}\BibitemShut {NoStop}%
\bibitem [{\citenamefont {Chern}\ \emph
  {et~al.}(2017{\natexlab{a}})\citenamefont {Chern}, \citenamefont {Hwang},
  \citenamefont {Mizoguchi}, \citenamefont {Huh},\ and\ \citenamefont
  {Kim}}]{Chern2017a}%
  \BibitemOpen
  \bibfield  {author} {\bibinfo {author} {\bibfnamefont {L.~E.}\ \bibnamefont
  {Chern}}, \bibinfo {author} {\bibfnamefont {K.}~\bibnamefont {Hwang}},
  \bibinfo {author} {\bibfnamefont {T.}~\bibnamefont {Mizoguchi}}, \bibinfo
  {author} {\bibfnamefont {Y.}~\bibnamefont {Huh}}, \ and\ \bibinfo {author}
  {\bibfnamefont {Y.~B.}\ \bibnamefont {Kim}},\ }\href {\doibase
  10.1103/PhysRevB.96.035118} {\bibfield  {journal} {\bibinfo  {journal} {Phys.
  Rev. B}\ }\textbf {\bibinfo {volume} {96}},\ \bibinfo {pages} {035118}
  (\bibinfo {year} {2017}{\natexlab{a}})}\BibitemShut {NoStop}%
\bibitem [{\citenamefont {Chern}\ \emph
  {et~al.}(2017{\natexlab{b}})\citenamefont {Chern}, \citenamefont {Schaffer},
  \citenamefont {Sorn},\ and\ \citenamefont {Kim}}]{Chern2017b}%
  \BibitemOpen
  \bibfield  {author} {\bibinfo {author} {\bibfnamefont {L.~E.}\ \bibnamefont
  {Chern}}, \bibinfo {author} {\bibfnamefont {R.}~\bibnamefont {Schaffer}},
  \bibinfo {author} {\bibfnamefont {S.}~\bibnamefont {Sorn}}, \ and\ \bibinfo
  {author} {\bibfnamefont {Y.~B.}\ \bibnamefont {Kim}},\ }\href {\doibase
  10.1103/PhysRevB.96.165117} {\bibfield  {journal} {\bibinfo  {journal} {Phys.
  Rev. B}\ }\textbf {\bibinfo {volume} {96}},\ \bibinfo {pages} {165117}
  (\bibinfo {year} {2017}{\natexlab{b}})}\BibitemShut {NoStop}%
\bibitem [{\citenamefont {Aidoudi}\ \emph {et~al.}(2011)\citenamefont
  {Aidoudi}, \citenamefont {Aldous}, \citenamefont {Goff}, \citenamefont {Z.},
  \citenamefont {Attfield}, \citenamefont {Morris},\ and\ \citenamefont
  {Lightfoot}}]{Aidoudi2011}%
  \BibitemOpen
  \bibfield  {author} {\bibinfo {author} {\bibfnamefont {F.~H.}\ \bibnamefont
  {Aidoudi}}, \bibinfo {author} {\bibfnamefont {D.~W.}\ \bibnamefont {Aldous}},
  \bibinfo {author} {\bibfnamefont {R.~J.}\ \bibnamefont {Goff}}, \bibinfo
  {author} {\bibfnamefont {S.~M.}\ \bibnamefont {Z.}}, \bibinfo {author}
  {\bibfnamefont {J.~P.}\ \bibnamefont {Attfield}}, \bibinfo {author}
  {\bibfnamefont {R.~E.}\ \bibnamefont {Morris}}, \ and\ \bibinfo {author}
  {\bibfnamefont {P.}~\bibnamefont {Lightfoot}},\ }\href
  {http://dx.doi.org/10.1038/nchem.1129} {\bibfield  {journal} {\bibinfo
  {journal} {Nat. Chem.}\ }\textbf {\bibinfo {volume} {3}},\ \bibinfo {pages}
  {801} (\bibinfo {year} {2011})}\BibitemShut {NoStop}%
\bibitem [{\citenamefont {Okamoto}\ \emph {et~al.}(2013)\citenamefont
  {Okamoto}, \citenamefont {Nilsen}, \citenamefont {Attfield},\ and\
  \citenamefont {Hiroi}}]{Okamoto2013}%
  \BibitemOpen
  \bibfield  {author} {\bibinfo {author} {\bibfnamefont {Y.}~\bibnamefont
  {Okamoto}}, \bibinfo {author} {\bibfnamefont {G.~J.}\ \bibnamefont {Nilsen}},
  \bibinfo {author} {\bibfnamefont {J.~P.}\ \bibnamefont {Attfield}}, \ and\
  \bibinfo {author} {\bibfnamefont {Z.}~\bibnamefont {Hiroi}},\ }\href
  {\doibase 10.1103/PhysRevLett.110.097203} {\bibfield  {journal} {\bibinfo
  {journal} {Phys. Rev. Lett.}\ }\textbf {\bibinfo {volume} {110}},\ \bibinfo
  {pages} {097203} (\bibinfo {year} {2013})}\BibitemShut {NoStop}%
\bibitem [{\citenamefont {Essafi}\ \emph {et~al.}(2017)\citenamefont {Essafi},
  \citenamefont {Jaubert},\ and\ \citenamefont {Udagawa}}]{Essafi2017}%
  \BibitemOpen
  \bibfield  {author} {\bibinfo {author} {\bibfnamefont {K.}~\bibnamefont
  {Essafi}}, \bibinfo {author} {\bibfnamefont {L.~D.~C.}\ \bibnamefont
  {Jaubert}}, \ and\ \bibinfo {author} {\bibfnamefont {M.}~\bibnamefont
  {Udagawa}},\ }\href {http://stacks.iop.org/0953-8984/29/i=31/a=315802}
  {\bibfield  {journal} {\bibinfo  {journal} {J. Phys.: Condens. Matter}\
  }\textbf {\bibinfo {volume} {29}},\ \bibinfo {pages} {315802} (\bibinfo
  {year} {2017})}\BibitemShut {NoStop}%
\bibitem [{\citenamefont {Mila}(1998)}]{Mila1998}%
  \BibitemOpen
  \bibfield  {author} {\bibinfo {author} {\bibfnamefont {F.}~\bibnamefont
  {Mila}},\ }\href {\doibase 10.1103/PhysRevLett.81.2356} {\bibfield  {journal}
  {\bibinfo  {journal} {Phys. Rev. Lett.}\ }\textbf {\bibinfo {volume} {81}},\
  \bibinfo {pages} {2356} (\bibinfo {year} {1998})}\BibitemShut {NoStop}%
\bibitem [{\citenamefont {Mambrini}\ and\ \citenamefont
  {Mila}(2000)}]{Mambrini2000}%
  \BibitemOpen
  \bibfield  {author} {\bibinfo {author} {\bibfnamefont {M.}~\bibnamefont
  {Mambrini}}\ and\ \bibinfo {author} {\bibfnamefont {F.}~\bibnamefont
  {Mila}},\ }\href {\doibase 10.1007/PL00011071} {\bibfield  {journal}
  {\bibinfo  {journal} {Eur. Phys. J. B}\ }\textbf {\bibinfo {volume} {17}},\
  \bibinfo {pages} {651} (\bibinfo {year} {2000})}\BibitemShut {NoStop}%
\bibitem [{\citenamefont {Orain}\ \emph {et~al.}(2017)\citenamefont {Orain},
  \citenamefont {Bernu}, \citenamefont {Mendels}, \citenamefont {Clark},
  \citenamefont {Aidoudi}, \citenamefont {Lightfoot}, \citenamefont {Morris},\
  and\ \citenamefont {Bert}}]{Orain2017}%
  \BibitemOpen
  \bibfield  {author} {\bibinfo {author} {\bibfnamefont {J.-C.}\ \bibnamefont
  {Orain}}, \bibinfo {author} {\bibfnamefont {B.}~\bibnamefont {Bernu}},
  \bibinfo {author} {\bibfnamefont {P.}~\bibnamefont {Mendels}}, \bibinfo
  {author} {\bibfnamefont {L.}~\bibnamefont {Clark}}, \bibinfo {author}
  {\bibfnamefont {F.~H.}\ \bibnamefont {Aidoudi}}, \bibinfo {author}
  {\bibfnamefont {P.}~\bibnamefont {Lightfoot}}, \bibinfo {author}
  {\bibfnamefont {R.~E.}\ \bibnamefont {Morris}}, \ and\ \bibinfo {author}
  {\bibfnamefont {F.}~\bibnamefont {Bert}},\ }\href {\doibase
  10.1103/PhysRevLett.118.237203} {\bibfield  {journal} {\bibinfo  {journal}
  {Phys. Rev. Lett.}\ }\textbf {\bibinfo {volume} {118}},\ \bibinfo {pages}
  {237203} (\bibinfo {year} {2017})}\BibitemShut {NoStop}%
\bibitem [{\citenamefont {Orain}\ \emph {et~al.}(2014)\citenamefont {Orain},
  \citenamefont {Clark}, \citenamefont {Bert}, \citenamefont {Mendels},
  \citenamefont {Attfield}, \citenamefont {Aidoudi}, \citenamefont {Morris},
  \citenamefont {Lightfoot}, \citenamefont {Amato},\ and\ \citenamefont
  {Baines}}]{Orain2014}%
  \BibitemOpen
  \bibfield  {author} {\bibinfo {author} {\bibfnamefont {J.~C.}\ \bibnamefont
  {Orain}}, \bibinfo {author} {\bibfnamefont {L.}~\bibnamefont {Clark}},
  \bibinfo {author} {\bibfnamefont {F.}~\bibnamefont {Bert}}, \bibinfo {author}
  {\bibfnamefont {P.}~\bibnamefont {Mendels}}, \bibinfo {author} {\bibfnamefont
  {P.}~\bibnamefont {Attfield}}, \bibinfo {author} {\bibfnamefont {F.~H.}\
  \bibnamefont {Aidoudi}}, \bibinfo {author} {\bibfnamefont {R.~E.}\
  \bibnamefont {Morris}}, \bibinfo {author} {\bibfnamefont {P.}~\bibnamefont
  {Lightfoot}}, \bibinfo {author} {\bibfnamefont {A.}~\bibnamefont {Amato}}, \
  and\ \bibinfo {author} {\bibfnamefont {C.}~\bibnamefont {Baines}},\ }\href
  {http://stacks.iop.org/1742-6596/551/i=1/a=012004} {\bibfield  {journal}
  {\bibinfo  {journal} {J. Phys.: Conf. Ser.}\ }\textbf {\bibinfo {volume}
  {551}},\ \bibinfo {pages} {012004} (\bibinfo {year} {2014})}\BibitemShut
  {NoStop}%
\bibitem [{\citenamefont {Clark}\ \emph {et~al.}(2013)\citenamefont {Clark},
  \citenamefont {Orain}, \citenamefont {Bert}, \citenamefont {De~Vries},
  \citenamefont {Aidoudi}, \citenamefont {Morris}, \citenamefont {Lightfoot},
  \citenamefont {Lord}, \citenamefont {Telling}, \citenamefont {Bonville},
  \citenamefont {Attfield}, \citenamefont {Mendels},\ and\ \citenamefont
  {Harrison}}]{Clark2013}%
  \BibitemOpen
  \bibfield  {author} {\bibinfo {author} {\bibfnamefont {L.}~\bibnamefont
  {Clark}}, \bibinfo {author} {\bibfnamefont {J.~C.}\ \bibnamefont {Orain}},
  \bibinfo {author} {\bibfnamefont {F.}~\bibnamefont {Bert}}, \bibinfo {author}
  {\bibfnamefont {M.~A.}\ \bibnamefont {De~Vries}}, \bibinfo {author}
  {\bibfnamefont {F.~H.}\ \bibnamefont {Aidoudi}}, \bibinfo {author}
  {\bibfnamefont {R.~E.}\ \bibnamefont {Morris}}, \bibinfo {author}
  {\bibfnamefont {P.}~\bibnamefont {Lightfoot}}, \bibinfo {author}
  {\bibfnamefont {J.~S.}\ \bibnamefont {Lord}}, \bibinfo {author}
  {\bibfnamefont {M.~T.~F.}\ \bibnamefont {Telling}}, \bibinfo {author}
  {\bibfnamefont {P.}~\bibnamefont {Bonville}}, \bibinfo {author}
  {\bibfnamefont {J.~P.}\ \bibnamefont {Attfield}}, \bibinfo {author}
  {\bibfnamefont {P.}~\bibnamefont {Mendels}}, \ and\ \bibinfo {author}
  {\bibfnamefont {A.}~\bibnamefont {Harrison}},\ }\href {\doibase
  10.1103/PhysRevLett.110.207208} {\bibfield  {journal} {\bibinfo  {journal}
  {Phys. Rev. Lett.}\ }\textbf {\bibinfo {volume} {110}},\ \bibinfo {pages}
  {207208} (\bibinfo {year} {2013})}\BibitemShut {NoStop}%
\bibitem [{\citenamefont {Schaffer}\ \emph {et~al.}(2017)\citenamefont
  {Schaffer}, \citenamefont {Huh}, \citenamefont {Hwang},\ and\ \citenamefont
  {Kim}}]{Schaffer2017}%
  \BibitemOpen
  \bibfield  {author} {\bibinfo {author} {\bibfnamefont {R.}~\bibnamefont
  {Schaffer}}, \bibinfo {author} {\bibfnamefont {Y.}~\bibnamefont {Huh}},
  \bibinfo {author} {\bibfnamefont {K.}~\bibnamefont {Hwang}}, \ and\ \bibinfo
  {author} {\bibfnamefont {Y.~B.}\ \bibnamefont {Kim}},\ }\href {\doibase
  10.1103/PhysRevB.95.054410} {\bibfield  {journal} {\bibinfo  {journal} {Phys.
  Rev. B}\ }\textbf {\bibinfo {volume} {95}},\ \bibinfo {pages} {054410}
  (\bibinfo {year} {2017})}\BibitemShut {NoStop}%
\bibitem [{\citenamefont {Repellin}\ \emph {et~al.}(2017)\citenamefont
  {Repellin}, \citenamefont {He},\ and\ \citenamefont
  {Pollmann}}]{Repellin2017}%
  \BibitemOpen
  \bibfield  {author} {\bibinfo {author} {\bibfnamefont {C.}~\bibnamefont
  {Repellin}}, \bibinfo {author} {\bibfnamefont {Y.-C.}\ \bibnamefont {He}}, \
  and\ \bibinfo {author} {\bibfnamefont {F.}~\bibnamefont {Pollmann}},\ }\href
  {\doibase 10.1103/PhysRevB.96.205124} {\bibfield  {journal} {\bibinfo
  {journal} {Phys. Rev. B}\ }\textbf {\bibinfo {volume} {96}},\ \bibinfo
  {pages} {205124} (\bibinfo {year} {2017})}\BibitemShut {NoStop}%
\bibitem [{\citenamefont {Abrikosov}(1965)}]{Abrikosov1965}%
  \BibitemOpen
  \bibfield  {author} {\bibinfo {author} {\bibfnamefont {A.~A.}\ \bibnamefont
  {Abrikosov}},\ }\href {\doibase 10.1103/PhysicsPhysiqueFizika.2.5} {\bibfield
   {journal} {\bibinfo  {journal} {Physics}\ }\textbf {\bibinfo {volume} {2}},\
  \bibinfo {pages} {5} (\bibinfo {year} {1965})}\BibitemShut {NoStop}%
\bibitem [{\citenamefont {Lu}\ \emph {et~al.}(2011)\citenamefont {Lu},
  \citenamefont {Ran},\ and\ \citenamefont {Lee}}]{Lu2011}%
  \BibitemOpen
  \bibfield  {author} {\bibinfo {author} {\bibfnamefont {Y.-M.}\ \bibnamefont
  {Lu}}, \bibinfo {author} {\bibfnamefont {Y.}~\bibnamefont {Ran}}, \ and\
  \bibinfo {author} {\bibfnamefont {P.~A.}\ \bibnamefont {Lee}},\ }\href
  {\doibase 10.1103/PhysRevB.83.224413} {\bibfield  {journal} {\bibinfo
  {journal} {Phys. Rev. B}\ }\textbf {\bibinfo {volume} {83}},\ \bibinfo
  {pages} {224413} (\bibinfo {year} {2011})}\BibitemShut {NoStop}%
\bibitem [{\citenamefont {Becca}\ and\ \citenamefont
  {Sorella}(2017)}]{BeccaBook}%
  \BibitemOpen
  \bibfield  {author} {\bibinfo {author} {\bibfnamefont {F.}~\bibnamefont
  {Becca}}\ and\ \bibinfo {author} {\bibfnamefont {S.}~\bibnamefont
  {Sorella}},\ }\href
  {https://www.cambridge.org/core/books/quantum-monte-carlo-approaches-for-correlated-systems/EB88C86BD9553A0738BDAE400D0B2900}
  {\emph {\bibinfo {title} {Quantum Monte Carlo Approaches for Correlated
  Systems}}}\ (\bibinfo  {publisher} {Cambridge University Press},\ \bibinfo
  {year} {2017})\BibitemShut {NoStop}%
\bibitem [{\citenamefont {Sorella}(2005)}]{Sorella2005}%
  \BibitemOpen
  \bibfield  {author} {\bibinfo {author} {\bibfnamefont {S.}~\bibnamefont
  {Sorella}},\ }\href {\doibase 10.1103/PhysRevB.71.241103} {\bibfield
  {journal} {\bibinfo  {journal} {Phys. Rev. B}\ }\textbf {\bibinfo {volume}
  {71}},\ \bibinfo {pages} {241103} (\bibinfo {year} {2005})}\BibitemShut
  {NoStop}%
\bibitem [{\citenamefont {Iqbal}\ \emph
  {et~al.}(2011{\natexlab{a}})\citenamefont {Iqbal}, \citenamefont {Becca},\
  and\ \citenamefont {Poilblanc}}]{Iqbal2011b}%
  \BibitemOpen
  \bibfield  {author} {\bibinfo {author} {\bibfnamefont {Y.}~\bibnamefont
  {Iqbal}}, \bibinfo {author} {\bibfnamefont {F.}~\bibnamefont {Becca}}, \ and\
  \bibinfo {author} {\bibfnamefont {D.}~\bibnamefont {Poilblanc}},\ }\href
  {\doibase 10.1103/PhysRevB.84.020407} {\bibfield  {journal} {\bibinfo
  {journal} {Phys. Rev. B}\ }\textbf {\bibinfo {volume} {84}},\ \bibinfo
  {pages} {020407} (\bibinfo {year} {2011}{\natexlab{a}})}\BibitemShut
  {NoStop}%
\bibitem [{\citenamefont {Becca}\ \emph {et~al.}(2015)\citenamefont {Becca},
  \citenamefont {Hu}, \citenamefont {Iqbal}, \citenamefont {Parola},
  \citenamefont {Poilblanc},\ and\ \citenamefont {Sorella}}]{Becca2015}%
  \BibitemOpen
  \bibfield  {author} {\bibinfo {author} {\bibfnamefont {F.}~\bibnamefont
  {Becca}}, \bibinfo {author} {\bibfnamefont {W.-J.}\ \bibnamefont {Hu}},
  \bibinfo {author} {\bibfnamefont {Y.}~\bibnamefont {Iqbal}}, \bibinfo
  {author} {\bibfnamefont {A.}~\bibnamefont {Parola}}, \bibinfo {author}
  {\bibfnamefont {D.}~\bibnamefont {Poilblanc}}, \ and\ \bibinfo {author}
  {\bibfnamefont {S.}~\bibnamefont {Sorella}},\ }\href
  {http://stacks.iop.org/1742-6596/640/i=1/a=012039} {\bibfield  {journal}
  {\bibinfo  {journal} {J. Phys.: Conf. Ser.}\ }\textbf {\bibinfo {volume}
  {640}},\ \bibinfo {pages} {012039} (\bibinfo {year} {2015})}\BibitemShut
  {NoStop}%
\bibitem [{ext()}]{extr}%
  \BibitemOpen
  \href@noop {} {}\bibinfo {note} {We have employed a quadratic fit with
  relative weighting by $1/X^2$, which weights the points at the left part of
  the graph more than points to the right. Hence, the nonlinear regression
  minimizes the quantity $\sum (Y_{\rm data} - Y_{\rm curve})^2 / X_{\rm
  curve}^2$ instead of the $\sum (Y_{\rm data} - Y_{\rm curve})^2$. The choice
  of the weight factor $1/X^2$ is determined by the fact that it minimizes the
  sum of the absolute values of the relative errors.}\BibitemShut {Stop}%
\bibitem [{\citenamefont {{Verstraete}}\ and\ \citenamefont
  {{Cirac}}(2004)}]{Verstraete2004b}%
  \BibitemOpen
  \bibfield  {author} {\bibinfo {author} {\bibfnamefont {F.}~\bibnamefont
  {{Verstraete}}}\ and\ \bibinfo {author} {\bibfnamefont {J.~I.}\ \bibnamefont
  {{Cirac}}},\ }\href@noop {} {\bibfield  {journal} {\bibinfo  {journal}
  {eprint arXiv:cond-mat/0407066}\ } (\bibinfo {year} {2004})},\ \Eprint
  {http://arxiv.org/abs/cond-mat/0407066} {cond-mat/0407066} \BibitemShut
  {NoStop}%
\bibitem [{\citenamefont {Schuch}\ \emph {et~al.}(2007)\citenamefont {Schuch},
  \citenamefont {Wolf}, \citenamefont {Verstraete},\ and\ \citenamefont
  {Cirac}}]{Schuch2007}%
  \BibitemOpen
  \bibfield  {author} {\bibinfo {author} {\bibfnamefont {N.}~\bibnamefont
  {Schuch}}, \bibinfo {author} {\bibfnamefont {M.~M.}\ \bibnamefont {Wolf}},
  \bibinfo {author} {\bibfnamefont {F.}~\bibnamefont {Verstraete}}, \ and\
  \bibinfo {author} {\bibfnamefont {J.~I.}\ \bibnamefont {Cirac}},\ }\href
  {\doibase 10.1103/PhysRevLett.98.140506} {\bibfield  {journal} {\bibinfo
  {journal} {Phys. Rev. Lett.}\ }\textbf {\bibinfo {volume} {98}},\ \bibinfo
  {pages} {140506} (\bibinfo {year} {2007})}\BibitemShut {NoStop}%
\bibitem [{\citenamefont {Schuch}\ \emph {et~al.}(2012)\citenamefont {Schuch},
  \citenamefont {Poilblanc}, \citenamefont {Cirac},\ and\ \citenamefont
  {P\'erez-Garc\'{\i}a}}]{Schuch2012}%
  \BibitemOpen
  \bibfield  {author} {\bibinfo {author} {\bibfnamefont {N.}~\bibnamefont
  {Schuch}}, \bibinfo {author} {\bibfnamefont {D.}~\bibnamefont {Poilblanc}},
  \bibinfo {author} {\bibfnamefont {J.~I.}\ \bibnamefont {Cirac}}, \ and\
  \bibinfo {author} {\bibfnamefont {D.}~\bibnamefont {P\'erez-Garc\'{\i}a}},\
  }\href {\doibase 10.1103/PhysRevB.86.115108} {\bibfield  {journal} {\bibinfo
  {journal} {Phys. Rev. B}\ }\textbf {\bibinfo {volume} {86}},\ \bibinfo
  {pages} {115108} (\bibinfo {year} {2012})}\BibitemShut {NoStop}%
\bibitem [{\citenamefont {Anderson}(1973)}]{Anderson1973}%
  \BibitemOpen
  \bibfield  {author} {\bibinfo {author} {\bibfnamefont {P.}~\bibnamefont
  {Anderson}},\ }\href {\doibase
  http://dx.doi.org/10.1016/0025-5408(73)90167-0} {\bibfield  {journal}
  {\bibinfo  {journal} {Mater. Res. Bull.}\ }\textbf {\bibinfo {volume} {8}},\
  \bibinfo {pages} {153 } (\bibinfo {year} {1973})}\BibitemShut {NoStop}%
\bibitem [{\citenamefont {Poilblanc}\ \emph {et~al.}(2012)\citenamefont
  {Poilblanc}, \citenamefont {Schuch}, \citenamefont {P\'erez-Garc\'{\i}a},\
  and\ \citenamefont {Cirac}}]{Poilblanc2012}%
  \BibitemOpen
  \bibfield  {author} {\bibinfo {author} {\bibfnamefont {D.}~\bibnamefont
  {Poilblanc}}, \bibinfo {author} {\bibfnamefont {N.}~\bibnamefont {Schuch}},
  \bibinfo {author} {\bibfnamefont {D.}~\bibnamefont {P\'erez-Garc\'{\i}a}}, \
  and\ \bibinfo {author} {\bibfnamefont {J.~I.}\ \bibnamefont {Cirac}},\ }\href
  {\doibase 10.1103/PhysRevB.86.014404} {\bibfield  {journal} {\bibinfo
  {journal} {Phys. Rev. B}\ }\textbf {\bibinfo {volume} {86}},\ \bibinfo
  {pages} {014404} (\bibinfo {year} {2012})}\BibitemShut {NoStop}%
\bibitem [{\citenamefont {Yang}\ and\ \citenamefont {Yao}(2012)}]{YangYao2012}%
  \BibitemOpen
  \bibfield  {author} {\bibinfo {author} {\bibfnamefont {F.}~\bibnamefont
  {Yang}}\ and\ \bibinfo {author} {\bibfnamefont {H.}~\bibnamefont {Yao}},\
  }\href {\doibase 10.1103/PhysRevLett.109.147209} {\bibfield  {journal}
  {\bibinfo  {journal} {Phys. Rev. Lett.}\ }\textbf {\bibinfo {volume} {109}},\
  \bibinfo {pages} {147209} (\bibinfo {year} {2012})}\BibitemShut {NoStop}%
\bibitem [{\citenamefont {Poilblanc}\ and\ \citenamefont
  {Schuch}(2013)}]{Poilblanc2013a}%
  \BibitemOpen
  \bibfield  {author} {\bibinfo {author} {\bibfnamefont {D.}~\bibnamefont
  {Poilblanc}}\ and\ \bibinfo {author} {\bibfnamefont {N.}~\bibnamefont
  {Schuch}},\ }\href {\doibase 10.1103/PhysRevB.87.140407} {\bibfield
  {journal} {\bibinfo  {journal} {Phys. Rev. B}\ }\textbf {\bibinfo {volume}
  {87}},\ \bibinfo {pages} {140407} (\bibinfo {year} {2013})}\BibitemShut
  {NoStop}%
\bibitem [{sup()}]{supp}%
  \BibitemOpen
  \href@noop {} {}\bibinfo {note} {See Supplemental Material at end of the main
  paper for the tables of energies of competing states}\BibitemShut {NoStop}%
\bibitem [{\citenamefont {{Iqbal}}\ \emph {et~al.}(2016)\citenamefont
  {{Iqbal}}, \citenamefont {{Poilblanc}},\ and\ \citenamefont
  {{Becca}}}]{Iqbal2016}%
  \BibitemOpen
  \bibfield  {author} {\bibinfo {author} {\bibfnamefont {Y.}~\bibnamefont
  {{Iqbal}}}, \bibinfo {author} {\bibfnamefont {D.}~\bibnamefont
  {{Poilblanc}}}, \ and\ \bibinfo {author} {\bibfnamefont {F.}~\bibnamefont
  {{Becca}}},\ }\href@noop {} {\bibfield  {journal} {\bibinfo  {journal} {ArXiv
  e-prints}\ } (\bibinfo {year} {2016})},\ \Eprint
  {http://arxiv.org/abs/1606.02255} {arXiv:1606.02255 [cond-mat.str-el]}
  \BibitemShut {NoStop}%
\bibitem [{\citenamefont {{Li}}(2016)}]{Li2016}%
  \BibitemOpen
  \bibfield  {author} {\bibinfo {author} {\bibfnamefont {T.}~\bibnamefont
  {{Li}}},\ }\href@noop {} {\bibfield  {journal} {\bibinfo  {journal} {ArXiv
  e-prints}\ } (\bibinfo {year} {2016})},\ \Eprint
  {http://arxiv.org/abs/1601.02165} {arXiv:1601.02165 [cond-mat.str-el]}
  \BibitemShut {NoStop}%
\bibitem [{\citenamefont {Iqbal}\ \emph {et~al.}(2012)\citenamefont {Iqbal},
  \citenamefont {Becca},\ and\ \citenamefont {Poilblanc}}]{Iqbal2012}%
  \BibitemOpen
  \bibfield  {author} {\bibinfo {author} {\bibfnamefont {Y.}~\bibnamefont
  {Iqbal}}, \bibinfo {author} {\bibfnamefont {F.}~\bibnamefont {Becca}}, \ and\
  \bibinfo {author} {\bibfnamefont {D.}~\bibnamefont {Poilblanc}},\ }\href
  {http://stacks.iop.org/1367-2630/14/i=11/a=115031} {\bibfield  {journal}
  {\bibinfo  {journal} {New J. Phys.}\ }\textbf {\bibinfo {volume} {14}},\
  \bibinfo {pages} {115031} (\bibinfo {year} {2012})}\BibitemShut {NoStop}%
\bibitem [{\citenamefont {Iqbal}\ \emph
  {et~al.}(2011{\natexlab{b}})\citenamefont {Iqbal}, \citenamefont {Becca},\
  and\ \citenamefont {Poilblanc}}]{Iqbal2011a}%
  \BibitemOpen
  \bibfield  {author} {\bibinfo {author} {\bibfnamefont {Y.}~\bibnamefont
  {Iqbal}}, \bibinfo {author} {\bibfnamefont {F.}~\bibnamefont {Becca}}, \ and\
  \bibinfo {author} {\bibfnamefont {D.}~\bibnamefont {Poilblanc}},\ }\href
  {\doibase 10.1103/PhysRevB.83.100404} {\bibfield  {journal} {\bibinfo
  {journal} {Phys. Rev. B}\ }\textbf {\bibinfo {volume} {83}},\ \bibinfo
  {pages} {100404} (\bibinfo {year} {2011}{\natexlab{b}})}\BibitemShut
  {NoStop}%
\end{thebibliography}
\renewcommand*{\citenumfont}[1]{S#1}
\renewcommand*{\bibnumfmt}[1]{[S#1]}
 
\newcommand\blankpage{%
    \null
    \thispagestyle{empty}%
    \addtocounter{page}{-1}%
    \newpage}

\blankpage

\chead{{\large \bf{\huge---\huge{Supplemental Material}---\\
}}}

\thispagestyle{fancy}

\beginsupplement

\begin{table}[ht]

\begin{minipage}{\textwidth}
\centering 
\begin{tabular}{lllllllll}
 \hline \hline
       \multicolumn{1}{c}{$J_{\triangledown}/J_{\vartriangle}$}
    & \multicolumn{1}{c}{$48$}
    & \multicolumn{1}{c}{$192$} 
    & \multicolumn{1}{c}{$432$}
    & \multicolumn{1}{c}{$768$}
    & \multicolumn{1}{c}{$1200$}
    & \multicolumn{1}{c}{$1768$}
    & \multicolumn{1}{c}{$2352$}  \\ \hline
    
\multirow{1}{*}{$1.0$} & $-0.4293926(15)$ & $-0.4287314(23)$ & $-0.4287114(3)$ & $-0.4287160(6)$ & $-0.4287168(27)$ & $-0.428717(7)$ & $-0.428708(9)$   \\    
       
\multirow{1}{*}{$0.9$} & $-0.4081210(5)$ & $-0.4074947(7)$ & $-0.4074783(7)$ & $-0.407480(1)$ & $-0.407487(2)$ & $-0.407492(3)$ & $-0.407492(4)$   \\ 
                                                                                                                       
\multirow{1}{*}{$0.7$} & $-0.3669648(5)$ & $-0.3664411(5)$ & $-0.3664274(7)$ & $-0.3664297(10)$ & $-0.3664373(15)$ & $-0.3664426(25)$ & $-0.3664442(35)$   \\ 
                                                                                                                                          
\multirow{1}{*}{$0.5$} & $-0.3282543(4)$ & $-0.3278732(4)$ & $-0.3278664(5)$ & $-0.3278714(9)$ & $-0.3278738(13)$ & $-0.327876(2)$ & $-0.327881(3)$    \\

\multirow{1}{*}{$0.4$} & $-0.3101170(3)$ & $-0.3098222(4)$ & $-0.3098184(9)$ & $-0.309822(2)$ & $-0.309826(3)$ & $$ & $$   \\ 

\multirow{1}{*}{$0.35$} & $-0.3014232(3)$ & $-0.3011736(4)$ & $-0.3011720(9)$ & $-0.3011782(16)$ & $-0.301181(3)$ & $$ & $$   \\ 

\multirow{1}{*}{$0.3$} & $-0.2930111(3)$ & $-0.2928055(3)$ & $-0.2928074(4)$ & $-0.2928122(6)$ & $-0.2928132(9)$ & $-0.2928167(15)$ & $-0.2928176(20)$   \\ 

\multirow{1}{*}{$0.25$} & $-0.2849048(2)$ & $-0.2847442(4)$ & $-0.2847466(10)$ & $-0.284751(2)$ & $-0.284760(3)$ & $$ & $$   \\

\multirow{1}{*}{$0.2$} & $-0.2771349(2)$ & $-0.2770175(3)$ & $-0.2770210(8)$ & $-0.2770233(15)$ & $-0.277021(2)$ & $$ & $$   \\

\multirow{1}{*}{$0.15$} & $-0.2697295(2)$ & $-0.2696521(3)$ & $-0.2696563(7)$ & $-0.2696576(13)$ & $-0.2696626(21)$ & $$ & $$   \\ 

\multirow{1}{*}{$0.1$} & $-0.26271957(11)$ & $-0.26267520(13)$ & $-0.26267882(13)$ & $-0.26268176(23)$ & $-0.2626831(4)$ & $-0.2626840(6)$ & $-0.2626847(8)$   \\ 

\multirow{1}{*}{$0.05$} & $-0.25613379(5)$ & $-0.25611611(8)$ & $-0.2561187(2)$ & $-0.2561202(3)$ & $-0.2561216(5)$ & $$ & $$   \\ \hline \hline

\end{tabular}
\caption{For various values of the breathing anisotropy $J_{\triangledown}/J_{\vartriangle}$, the variational ground-state energies per site $E/J_{\vartriangle}$ of the gapless $U(1)$ Dirac spin liquid on different cluster sizes (labelled by the total number of sites) is given. The $U(1)$ Dirac spin liquid Ansatz employed includes both the nearest-neighbor and optimized next-nearest-neighbor hopping. The calculations are done using mixed boundary conditions, i.e., anti-periodic along {\bf a}$_{1}$ and periodic along {\bf a}$_{2}$. All the clusters are of the type $3{\times}L{\times}L$, and do not explicitly break lattice symmetries.}
\label{tab:U1}
\end{minipage}\vspace{1cm}

\begin{minipage}{\textwidth}
\centering
\begin{tabular}{lllllllll}
 \hline \hline
       \multicolumn{1}{c}{$J_{\triangledown}/J_{\vartriangle}$}
    & \multicolumn{1}{c}{$48$}
    & \multicolumn{1}{c}{$192$} 
    & \multicolumn{1}{c}{$432$}
    & \multicolumn{1}{c}{$768$}
    & \multicolumn{1}{c}{$1200$}
    & \multicolumn{1}{c}{$1768$}
    & \multicolumn{1}{c}{$2352$}  \\ \hline
    
\multirow{1}{*}{$1.0$} & $-0.4295356(12)$ & $-0.4287638(23)$ & $-0.4287266(4)$ & $-0.4287204(7)$ & $-0.4287177(37)$ & $-0.428725(10)$ & $-0.428711(12)$   \\    
       
\multirow{1}{*}{$0.9$} & $-0.4082574(4)$ & $-0.4075283(5)$ & $-0.4074916(9)$ & $-0.4074900(12)$ & $-0.4074925(14)$ & $-0.407491(2)$ & $-0.407493(3)$   \\ 
                                                                                                                       
\multirow{1}{*}{$0.7$} & $-0.36708095(33)$ & $-0.3664724(5)$ & $-0.3664417(8)$ & $-0.3664408(11)$ & $-0.3664425(13)$ & $-0.366440(2)$ & $-0.366437(3)$   \\ 
                                                                                                                                          
\multirow{1}{*}{$0.5$} & $-0.32834733(26)$ & $-0.3279079(3)$ & $-0.3278864(7)$ & $-0.3278839(10)$ & $-0.3278842(10)$ & $-0.3278835(15)$ & $-0.3278794(26)$    \\

\multirow{1}{*}{$0.3$} & $-0.29309867(26)$ & $-0.2928537(3)$ & $-0.2928370(5)$ & $-0.2928321(7)$ & $-0.2928300(6)$ & $-0.2928278(11)$ & $-0.2928240(14)$   \\ 

\multirow{1}{*}{$0.1$} & $-0.26283227(10)$ & $-0.26275058(10)$ & $-0.2627398(3)$ & $-0.2627301(2)$ & $-0.2627241(5)$ & $-0.2627230(6)$ & $-0.2627147(7)$   \\ \hline \hline

\end{tabular}
\caption{For various values of the breathing anisotropy $J_{\triangledown}/J_{\vartriangle}$, the variational ground-state energies per site $E/J_{\vartriangle}$ of the gapped $\mathbb{Z}_{2}[0,\pi]\beta^{*}$ spin liquid on different cluster sizes (labelled by the total number of sites) is given. The calculations are done using mixed boundary conditions, i.e., anti-periodic along {\bf a}$_{1}$ and periodic along {\bf a}$_{2}$. All the clusters are of the type $3{\times}L{\times}L$, and do not explicitly break lattice symmetries.}
\label{tab:Z2}
\end{minipage}\vspace{1cm}

\begin{minipage}{\textwidth}
\begin{footnotesize}
\centering
\begin{tabular}{lllllllll}
 \hline \hline
       \multicolumn{1}{c}{Size}
    & \multicolumn{1}{c}{$0$-LS}
    & \multicolumn{1}{c}{$1$-LS} 
    & \multicolumn{1}{c}{$2$-LS}
    & \multicolumn{1}{c}{$0$-LS}
    & \multicolumn{1}{c}{$1$-LS}
    & \multicolumn{1}{c}{$2$-LS}
    & \multicolumn{1}{c}{$U(1)_{\sigma^{2}=0}$}  
    & \multicolumn{1}{c}{$\mathbb{Z}_{2}[0,\pi]\beta^{*}_{\sigma^{2}=0}$} \\ \hline
       
\multirow{1}{*}{$48$} & $-0.3282543(4)$ & $-0.33174706(30)$ & $-0.3328861(9)$ & $-0.32834733(26)$ & $-0.33175460(23)$ & $-0.3328805(9)$ & $-0.334835(30)$ & $-0.334702(27)$  \\ 
                                                                                                                       
\multirow{1}{*}{$192$} & $-0.3278732(4)$ & $-0.3307906(6)$ & $-0.331946(4)$ & $-0.3279079(3)$ & $-0.3308047(6)$ & $-0.331947(3)$ & $-0.334564(78)$ & $-0.334167(83)$  \\ \hline \hline                                                                                                                              
\end{tabular}
\caption{At $J_{\triangledown}/J_{\vartriangle}=0.5$, the variational ground-state energies of the $U(1)$ Dirac spin liquid (columns $2{-}4$) and the $\mathbb{Z}_{2}[0,\pi]\beta^{*}$ spin liquid (columns $5{-}7$), with $p=0$, $1$, and $2$ Lanczos steps on different cluster sizes obtained by VMC are given. The (non-variational) estimate of the ground-state energy of the $S=1/2$ Heisenberg model on different cluster sizes obtained from a zero-variance extrapolation (employing a quadratic fit) of the $0$, $1$, and $2$ Lanczos step energies of both the $U(1)$ and $\mathbb{Z}_{2}[0,\pi]\beta^{*}$ spin liquid Ans\"atze is given in columns $8{-}9$.}
\label{tab:en-lanczos}
\end{footnotesize}
\end{minipage}\vspace{1cm}

\begin{minipage}{\textwidth}
\centering
\begin{tabular}{lllllll}
 \hline \hline
       \multicolumn{1}{c}{$J_{\triangledown}/J_{\vartriangle}$}
    & \multicolumn{1}{c}{$48$}
    & \multicolumn{1}{c}{$192$} 
    & \multicolumn{1}{c}{$432$}
    & \multicolumn{1}{c}{$768$}
    & \multicolumn{1}{c}{$1200$}  \\ \hline    
       
\multirow{1}{*}{$0.05$} & $-0.25640035(4)$ & $-0.25640896(11)$ & $-0.2564084(5)$ & $-0.2564069(5)$ & $-0.2564057(7)$ \\        
       
\multirow{1}{*}{$0.1$} & $-0.26305787(9)$ & $-0.2630934(2)$ & $-0.2630934(6)$ & $-0.263081(2)$ & $-0.263088(1)$ \\ 

\multirow{1}{*}{$0.15$} & $-0.2699562(1)$ & $-0.2700349(3)$ & $-0.2700379(7)$ & $-0.270037(1)$ & $-0.270030(2)$  \\ 

\multirow{1}{*}{$0.2$} & $-0.2771352(2)$ & $-0.2772260(4)$ & $-0.277236(1)$ & $-0.277230(1)$ & $-0.277222(2)$  \\ \hline \hline
                                                                                                                                                                                                                                                             
\end{tabular}
\caption{For various values of the breathing anisotropy $J_{\triangledown}/J_{\vartriangle}$ inside the valance-bond crystal ordered phase, the variational ground-state energies per site $E/J_{\vartriangle}$ of the 6-site unit-cell valence-bond crystal on different cluster sizes (labelled by the total number of sites) is given. The valence-bond crystal is obtained by dimerizing the nearest-neighbor hopping amplitudes of the $U(1)$ Dirac spin liquid, whereas the second-nearest neighbor hoppings are not dimerized. The calculations are done using mixed boundary conditions, i.e., anti-periodic along {\bf a}$_{1}$ and periodic along {\bf a}$_{2}$. All the clusters are of the type $3{\times}L{\times}L$, and do not explicitly break lattice symmetries.}
\label{tab:vbc}
\end{minipage}

\end{table}

\end{document}